# A Deterministic Method to Calculate the AIS Trauma Score from a Finite Element Organ Trauma Model (OTM)

C. Bastien*, C. Neal-Sturgess**, J. Christensen*, L. Wen*

*Institute for Transport and Cities, Coventry University, Priory Street, Coventry, CV1 5FB, UK

** University of Birmingham, School of Mechanical Engineering, Edgbaston, Birmingham, B15 2TT, UK

**Abstract:** Traumatic injuries are measured using the Abbreviated Injury Scale (AIS), which is a risk to life scale. New human computer models use stresses and strains to evaluate whether serious or fatal injuries are reached, unfortunately these tensors bear no direct relation to AIS. This paper proposes to overcome this deficiency and suggests a unique Organ Trauma Model (OTM) able to calculate the risk to life based on the severity on any organ injury, focussing on real-life pedestrian accidents. The OTM uses a power method, named Peak Virtual Power (PVP), and calculates the risk to life of brain white and grey matters as a function of impact direction and impact speed. The OTM firstly calibrates PVP against the medical critical AIS threshold observed in each part of the head as a function of speed. This base PVP critical trauma function is then scaled and banded across all AIS levels using the confirmed property that AIS and the probability of death is statistically and numerically a cubic one. The OTM model has been tested against four real-life pedestrian accidents and proven to be able to predict pedestrian head trauma severity. In some cases, the method did however under-estimate the head trauma by 1 AIS level, because of post-impact haemorrhage which cannot be captured with the employed Lagrangian Finite Element (FE) solver. It is also shown that the location of the injury predictions using PVP coincide with the post mortem reports and are different to the predictions made using maximum principal strain.

# 1.0 Introduction

## 1.1 State of the Art injury indicators

Automotive Manufacturers design vehicles against legislative and consumer test protocols using crash test dummies with the purpose of creating safer vehicles for occupants and pedestrians. In spite of all their efforts, the number of fatalities keeps on increasing worldwide year by year [1], reaching 1.35 million in 2018. There are many parameters which can be attributed to this increase of death toll such as age, gender, speeding, etc., however, the steady rise in numbers begs the question whether the design tools currently used in the design process namely crash test dummies, are adequate to reverse this trend. Crash test dummies are anthropometric mechanical systems which can capture displacements, accelerations and forces, but do not contain internal organs. During the vehicle design process, dummies output information during the crash event which is cross-correlated to a probability of threat to life, based on injury severity. This trauma injury severity has been defined by medical professionals who have suggested a trauma injury scale or the Abbreviated Injury Scale (AIS). The AIS is an anatomically based, consensus derived, global severity scoring system that classifies each injury by body region according to its relative importance (threat to life) on a 6-point ordinal scale [2]. The latest revision of the AIS scale, which dates to 2015 [3], provides a standardised terminology to describe injuries and ranks injuries by severity. AIS is internationally accepted and is the primary tool to conclude injury severity [2]. From an engineering perspective, injury can be estimated using engineering indicators based on injury criteria. Currently, injury indicators can be classified into two major categories: kinematics-based indicators, used in crash test dummies; and strain-based indicators, when using a human computer model.

The first category relates to kinematics-based criteria which describe the kinematic behavior of a structure. However, such criteria, for example the Head Injury Criterion (HIC), cannot be used to describe the material response during impact [4]. The second category relates to human computer models, like THUMS [6] [11][12], which contain internal organs. In this case, plastic strains are used as the criterion in bone fracture and principal strain is often used in organ injury studies. In the case of the THUMS human model [6][11][12] plastic strain criteria is used to evaluate the maximum AIS, with the following threshold listed in Table 1.

| Tissue/Organ | Currently-used injury measurement | Injury description | AIS |
|---|---|---|---|
| Brain grey matter | Maximum 30% principal strain | Brain contusion | 3-4 |
| Brain white matter | 21% maximum principal strain | Diffuse Axonal Injury (DAI) | 4 |
| Heart | 30% maximum principal strain | Rupture | 4 |
| Liver | 30% maximum principal strain | Rupture | 4 |
| Spleen | 30% maximum principal strain | Rupture | 4 |
| Kidneys | 30% maximum principal strain | Rupture | 4 |
| Skull | Maximum 3% plastic strain | Fracture | 2-3 |

Table 1 currently used injury criterion on brain and organs and corresponding AIS level [12]

However, in human body injury, elastic strain decreases when the load is decreasing, whereas plastic strain remains; and on a human body, injury remains although the impact pulse is removed. Therefore, in concept, elastic strain-based indicators are different to human body injury. Also, time effects, or strain rates, are not considered when using strain as an injury indicator. Considering the Eiband injury graphs [7], injuries are linked with impulse duration, hence considering a time dependency factor when computing trauma. Currently (2019), the trauma location cannot be predicted using the strain-based method. As a consequence, it can be concluded that kinematic and strain-based indicators are not realistic metrics to assess trauma injury. Also, when Table 1 is used in trauma assessment, it is not possible to conclude which plastic strain level represents AIS 1, 2, 3, 4 and 5.

Such limitations can be overcome using the Peak Virtual Power method (PVP) which was shown to statistically correlate with trauma observed in real-life accidents [8][9][10]. PVP is an energy-based engineering indicator which was proposed as an injury criteria, and is derived from the rate dependent form of the 2nd law of thermodynamics using the Clausius-Duhem inequality, considering that irreversible work in a human body is equivalent to injury [8][9][10]. PVP takes the peak value of virtual power which indicates that it is monotonically increasing throughout the time history of an impact (Figure 1), and has been statistically proven to correlate with injury severity, with correlation coefficients ($R^2$) better than 0.98 [8][9][10], yet it has never been applied in a Finite Element formulation

On organ/tissue level, PVP can be extracted using the formula from Equation 1[8][9][10]:

$$PVP \propto AIS \propto max(\sigma \cdot \dot{\varepsilon})$$

Equation 1: Peak Virtual Power (PVP) formulation

Following Equation 1, PVP is extracted by multiplying the stress $\sigma$ and the strain rate $\dot{\varepsilon}$ and memorizing the maximum value as the impact event is taking place, as illustrated in Figure 1. Trauma, the maximum value of the PVP, remains present during the duration of the impact and does not reduce when the load is removed.

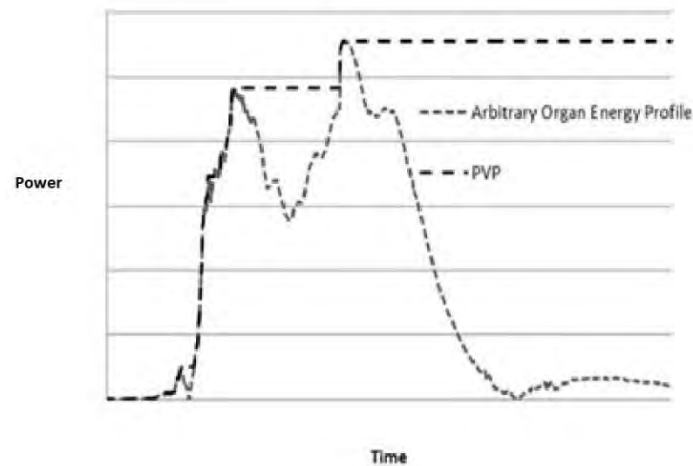

Figure 1: Illustartion of the PVP Concept

In order to extract PVP, several tensor candidates are available and need evaluation:

1. Plastic (stress/strain) component indicates that the stress/strain occurring in the plastic stage of the material when the yield stress is exceeded. This indicator is not adequate, because internal organs/tissues are made of water and collagen which are modelled as incompressible viscoelastic materials [8][9][10]. Under high strain-rate deformation, viscoelastic materials can essentially behave as elastic, which means that when using these material models there is no plastic components available from which to calculate PVP.

2. Principal (stress/strain) is the component acting on the main or principal plane where the shear is zero. Biomechanical injuries are the result of the separation (fracture, shearing, tearing or rupture) of biological tissues, hence ignoring shear components is also not representative.
3. The Von Mises criterion is the vector resultant from maximum shear stress, although it is usually quoted in terms of principal stresses. It is usually used as yield criterion for plastic deformation, taking shear into account.

Consequently, PVP, and therefore the AIS trauma level, can be extracted using the Von Mises vector resultant. The research proposed will answer the question whether trauma injury, extracted using PVP and then coded into an AIS, can be directly and deterministically extracted from a finite element model. The next section will discuss which parameters within the PVP equation affect trauma.

**1.2 Physical parameters influencing PVP**

When the PVP theory was derived from first principles using the Clausius-Duhem inequality, it was proven to accurately predict trauma against statistical real-life accident scenarios [8][9][10]. The base PVP theory fully correlated with belted and unbelted occupants accident data, and suggested that their respective trauma injury was a function of a cubic for belted and a square of the impact velocity for unbelted occupants, as per Equation 2 and Equation 3 [8][9][10].

$$PVP \propto AIS \propto V^2 \text{ [unbelted occupants]}$$

Equation 2: Relationship between PVP and velocity for unbelted occupants and belted occupants

$$PVP \propto AIS \propto V^3 \text{ [belted occupants]}$$

Equation 3: Relationship between PVP and velocity for belted occupants

In the case of pedestrian accidents (Figure 2), the real-life pedestrian accident data demonstrated that statistically the pedestrian trauma to impact velocity was proportional to the square of the impact speed for slight injuries and to the cubic of the impact speed for serious/ fatal injuries; but for the pedestrian cases no PVP theoretical derivations were successfully achieved to correlate with the real-life accident data.

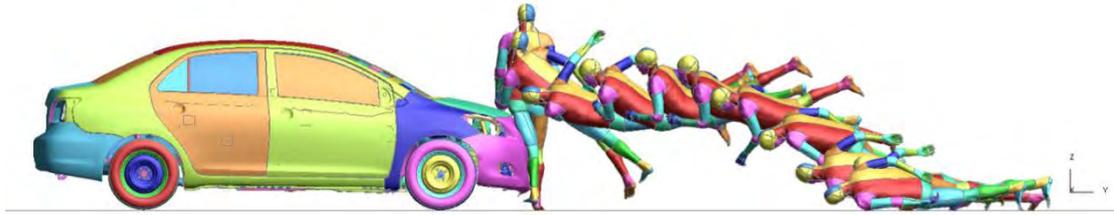

Figure 2: A typical pedestrian accident – Pedestrian kinematics

This section of the paper will answer two main points:
1. The first point is to understand the parameters which influence PVP, i.e. trauma injury. As the first derivation of PVP was using Continuum Damage Mechanics thermodynamic principles, another derivation will be here performed to highlight the key variable affecting trauma. This will be relevant later on in the paper.
2. The final point is to derive a theoretical PVP equation to confirm the trauma cubic relationship in the case of pedestrian impact.

During an impact, the kinetic energy of the organ is converted into strain energy, which is highlighted in Equation 4.

Organ kinetic energy = Organ strain energy

Equation 4: Conversion of Energy from Kinetic into strain during the impact

This energy transfer can be written mathematically as Equation 5, where m is the mass of the organ, v the impact speed, E the organ Young's modulus, and $\sigma$ the von Mises stress.

$$\frac{1}{2}mv^2 = \frac{\sigma^2}{2E}vol$$

Equation 5 Transfer of Energy from Kinetic into strain during the impact (re-formulation)

Using the fact that m=$\rho \cdot vol$ (ρ is the density, 'vol' is the volume of tissue/organ), this leads to Equation 6, which represents the link between stress and velocity.

$$\sigma = \sqrt{\rho E}v$$

Equation 6: Relationship between stress and velocity

Utilising Equation 1 and Equation 6 as well as considering the strain rate $\dot{\varepsilon}$, PVP can be re-written as algebraic transformations illustrated in Equation 7.

$$\dot{\varepsilon} = \frac{v}{L}$$

$$PVP = \frac{1}{L}\sqrt{\rho E}v^2$$

Equation 7: Final Derivation of PVP as a function of geometry and material properties

It can be observed from Equation 7 that PVP depends on the organ material property (E and ρ) and its size/ shape (L). Consequently, it can be concluded that PVP is direction dependant, i.e. that the trauma injury sustained will depend on the impact direction. These statements answer the first point of this paper, which was to capture the parameters influencing trauma injury.

In the case of pedestrian trauma relationship with impact speed, in the case of serious injuries, following the statistical fits for occupants, it can be hypothesised that the impact energy needs to include the pedestrian ride down on the bonnet. Indeed, if the body is in contact with the vehicle for a longer duration, the ride-down needs also to be considered. For the shorter contact times, ride-down can be ignored, because pedestrian and vehicle separate very quickly, so the ride-down does not have any effect. As PVP is power based, it can be assumed that it is proportional to the rate of impact energy, as illustrated in Equation 10.

$$Impact\ Kinetic\ Energy \propto v^2 \rightarrow PVP \propto \frac{v^2}{\Delta t}$$

Equation 8: Effect of ride-down in pedestrian scenarios

Assuming that the pedestrian impact velocity reduces linearly during the impact, its ride-down (S) can be expressed as Equation 11 (Newton second law)

$$S = \frac{v}{2}\Delta t \ \ or\ \ \Delta t = \frac{2S}{v}$$

Equation 9 Vehicle and pedestrian ride down – coupled system (crush distance)

By combining Equation 11 back into Equation 10, it can be shown that in the case of pedestrians, for serious injuries, the relationship between PVP and impact velocity is a cubic (Equation 12), as observed statistically in the real-life accident scenarios. This fact will be used later on in the paper.

$$PVP \propto Kv^3$$

Equation 10 Relationship between PVP and impact velocity in an uncoupled (pedestrian) impact

### 1.3 Purpose of the research

The proposed research aims at answering the question whether it is possible to extract the injury severity for soft tissue AIS organ injuries directly from the finite element model. The introduction section has highlighted that the trauma level, which can be calculated via PVP, was material, geometry, velocity and impact direction dependent. Using these four characteristics, it is proposed to create corridors of survivability (as a function of impact speed) and to test them against real-life scenarios in order to validate whether the PVP method is suitable to predict the trauma location as well as the trauma severity in a finite element environment. This study will be conducted using pedestrian accidents for which Police accident reports and Post-Mortems (PM) have been made available from the UK Police Force (UKPF). As most pedestrians die of head injuries [5], this paper will focus on defining a mechanical indicator to calculate the risk to life on brain tissues. The proposed research will be conducted in accordance with the Coventry University [17] and the NHS ethics protocols, ensuring respect of the deceased and full anonymity of data. An Information Sharing Agreement (ISA) has been signed between the UKPF and Coventry University setting the ethical and procedural requirements which have been met [18].

## 2.0  Methodology

The methodology used in this study is based on two phases. The first phase is the definition of the organ traumatology model (OTM) and the second one the traumatology model validation, based on real-world accident reconstruction; the phase I OTM method is pictured in Figure 2.

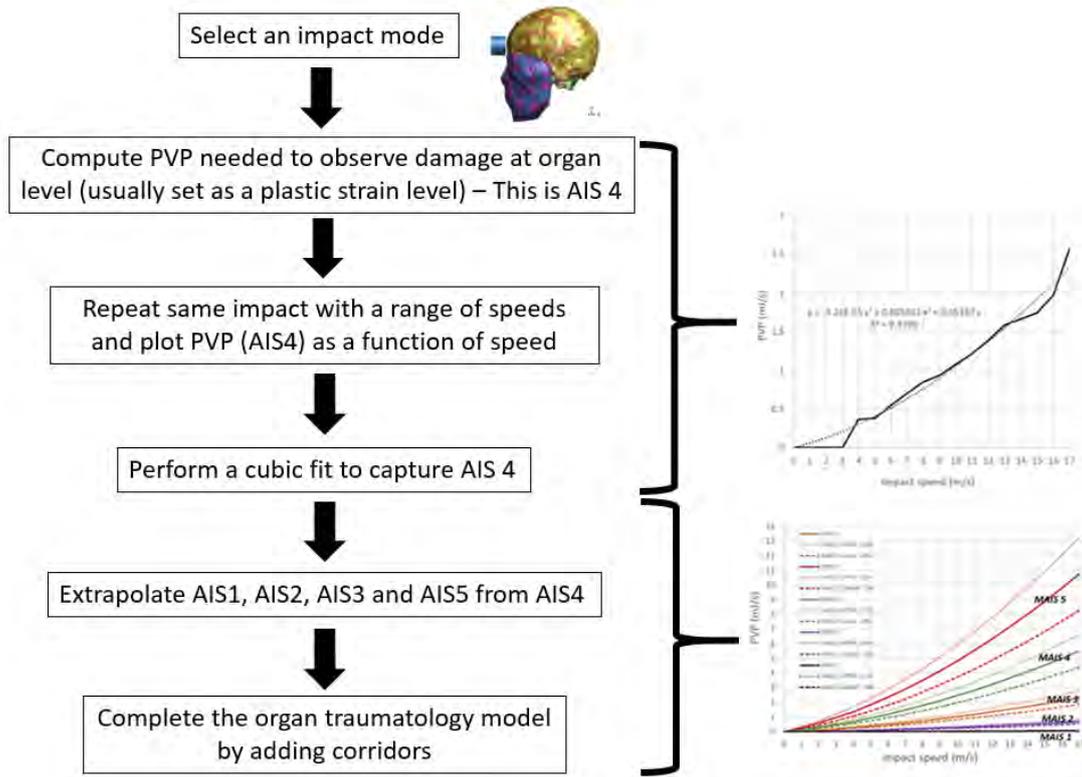

Figure 3: PHASE I: Organ Traumatology Model (OTM)

The research will consider the traumatology of the head in pedestrian impact scenarios and focus on the white and grey matter. The process starts by impacting the head in the three strategic locations (frontal, lateral and occipital), as documented by previous research [14] of pedestrian impacts, at velocity increments ranging from 2m/s to 17m/s. The upper value of 17m/s is the maximum velocity observed in the accidents provided by the UKPF. Also, this velocity relates to the maximum 64km/h frontal impact. Consequently, it is proposed that the range of speeds proposed would allow the use of this OTM in other modes of impact scenarios.

For each impact velocity in a defined scenario (frontal, lateral and occipital), the PVP of the first element in an organ reaching the critical level listed in Table 1, known as the threshold critical calibrated AIS value; the PVP value extracted at the time the damage is observed and plotted as a function of speed. This plot represents AIS 4 of this organ. The next stage is to capture the intermediate and ultimate AIS levels (AIS1, AIS2, AIS3 and AIS5), as well as their level of uncertainty.

Various studies collecting previous clinical research [8] have recorded the relationship between AIS and the risk to life. This data is plotted in Figure 3, and contains data from Baker, CCIS, NASS and Walder [13]. In order to remove the bias from each of the studies, the results from all the studies were averaged and extrapolated with a cubic relationship as well as including a 95% confidence level corridor, as illustrated in Figure 6. It was previously observed that the risk to life and the probability of death were related to a cubic ($R^2 > 0.95$) [8][9][10].

At this point it is important to note that the cubic fit does not aim at interpolating between the AIS values, which are ordinal values; the interpolation function is only interrogated at integer AIS levels. The cubic relationship confirms that at the ordinal AIS values, the relationship between trauma levels is a cubic in the "frequency of death".

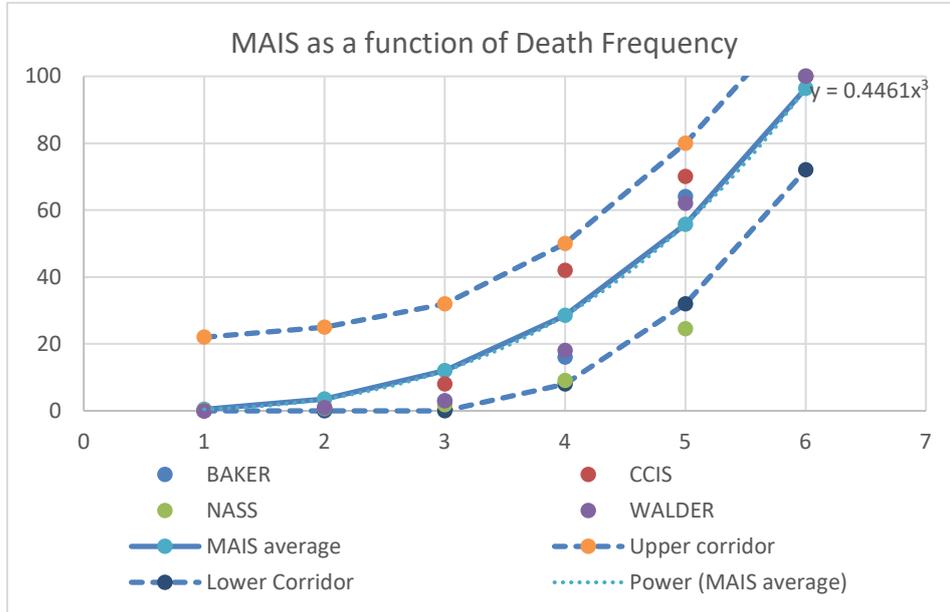

Figure 4 Curve fitting of MAIS and Probability of Fatality

Consequently, the probability of fatality of MAIS 5 can be expressed as Equation 12.

$$\frac{Probablity\ of\ fatality\ of\ MAIS5}{Probablity\ of\ fatality\ of\ MAIS4} = \frac{MAIS5^3}{MAIS4^3} = \frac{125}{64}$$

Equation 11: probability of fatality of MAIS 5

Hence the ratio of AIS3, AIS2 and AIS1 taking AIS4 as reference are 27/64, 8/64 and 1/64 respectively.

The AIS tolerance corridors, based on the clinical studies used, can be extracted from Figure 3 and are listed in Table 2 and concludes the OTM trauma model generation (PHASE I).

| MAIS level | Tolerance bound |
|---|---|
| 1 | +/- 21% |
| 2 | +/- 20% |
| 3 | +/- 20% |
| 4 | +/- 20% |
| 5 | +/- 23% |

Table 2 Tolerance bounds of each MAIS level

Phase II will aim at validating the OTM model (Figure 4).

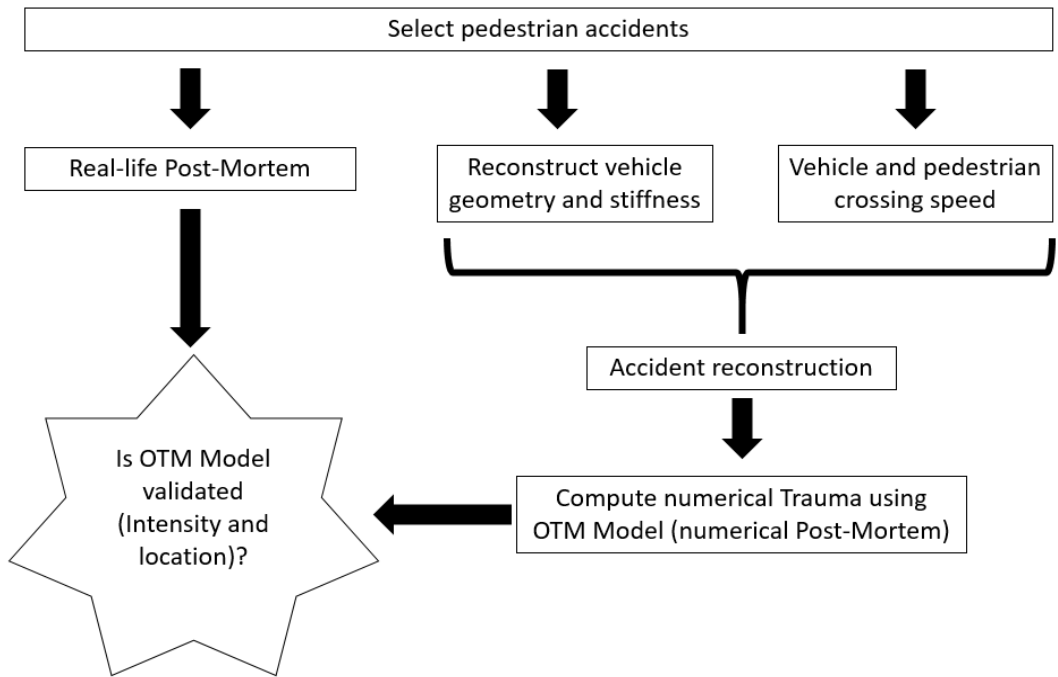

Figure 5: PHASE II. Validation of OTM trauma model

To do so, four accidents have been provided by the UKPF. For each accident, the real-life trauma is extracted from the Post-Mortem provided by the Coroner. The accident is numerically reconstructed, and the numerical trauma computed using the OTM model from Phase I.

In order to validate the OTM method, the trauma intensity (AIS level) and trauma location will be compared to the real-life trauma extracted from the post-mortem (PM).

## 3.0 Phase I: Calibration of OTM Trauma Model

A cylindrical impactor of 200g was created and positioned around the THUMS human head computer model in the forehead (Figure 6), lateral (Appendix A) and occipital (Appendix B) areas,. This approach was selected because the impact severity depends on the impact location [14]. The frontal impact computer model is illustrated in Figure 5. The temporal and occipital impact models and interpolations can be found in Appendix A and Appendix B respectively.

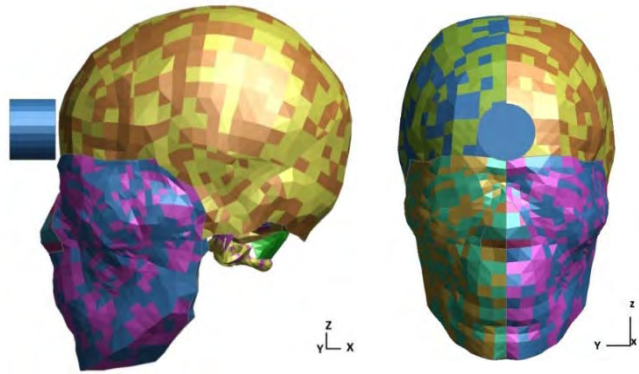

Figure 6 Scenario of frontal impact on THUMS' head

As described in the methodology, the impacts were conducted from 2.0m/s to 17.0m/s in 1m/s increments. Once the threshold plastic strain injury criterion is reached in one of the elements of the head, its PVP value at that specific time is extracted and plotted against impact speed. Brain contusion (grey matter) and Diffuse Axon Injury (DAI) (white matter) are classified as AIS 3 and AIS 4 level injuries respectively. Consequently, the PVP obtained for the critical plastic strain threshold obtained on grey matter relates to an AIS 3 brain contusion. The PVP threshold obtained on the brain white matter is equivalent to an AIS 4 Diffuse Axonal Injury (DAI). In the frontal impact scenario, the PVP threshold of AIS 3 brain contusion and AIS 4 DAI are shown in Figure 6 and Figure 7 respectively, including the corridors in Table 2.

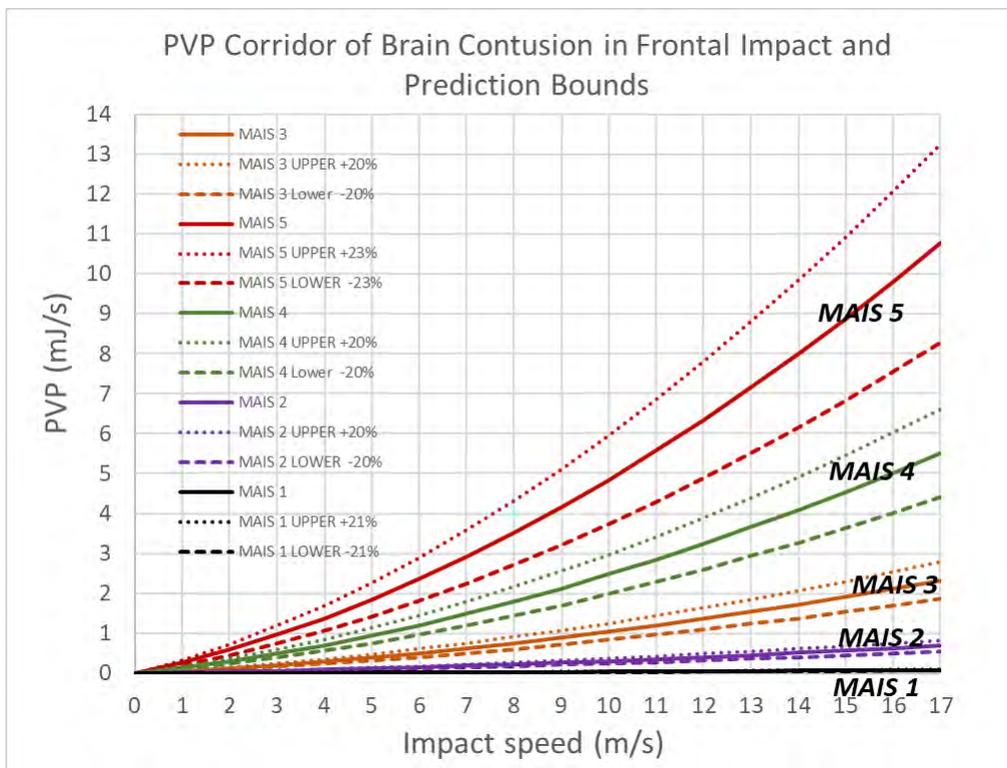

Figure 7: PVP corridor of brain contusion in frontal impact (grey matter)

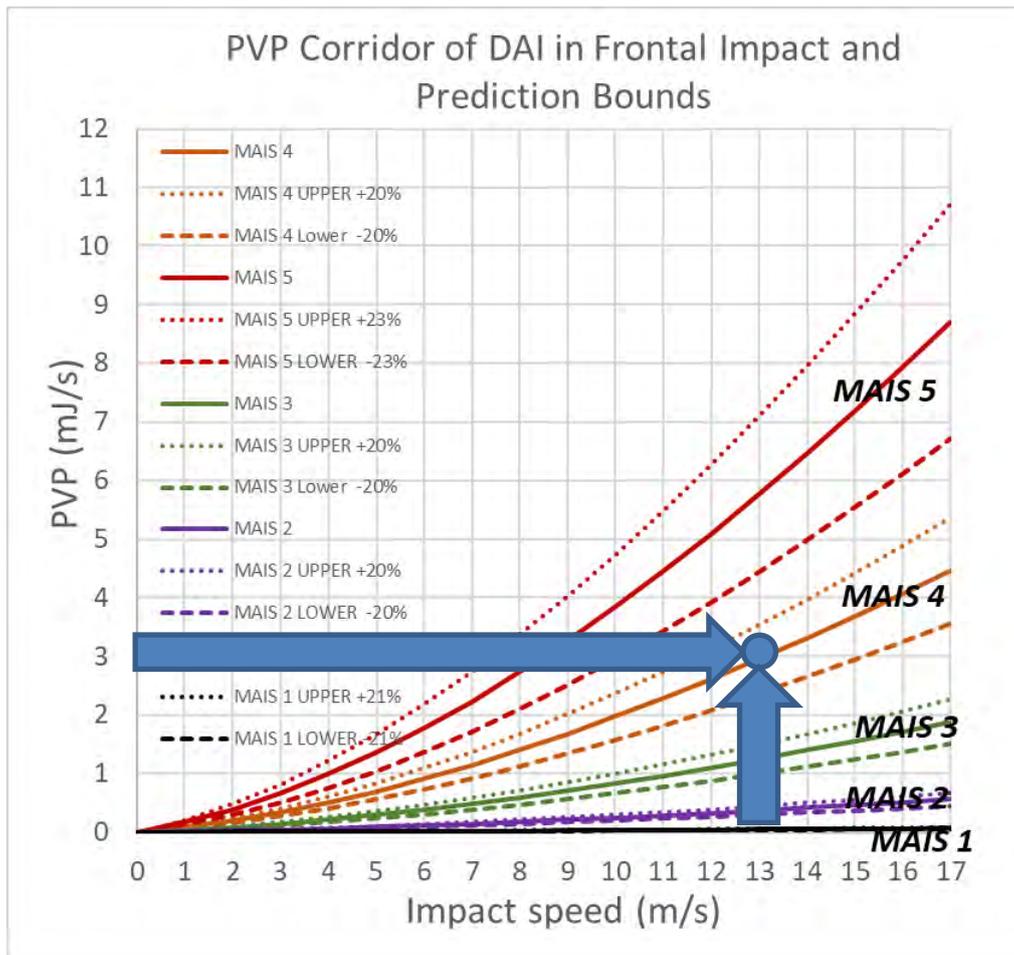

Figure 8: PVP corridor of DAI in frontal impact (white matter)

The graphs plotted, Figure 6 and Figure 7, represent an Organ Trauma Model (OTM), which maps the whole trauma response of an organ in a specific impact direction, against an impact speed. For a selected impact speed (abscise), the PVP can be read from the finite element model (ordinate). The AIS level can be extracted from these two values. As an example, in Figure 7, for an impact speed of 13m/s, should the PVP read 3mJ/s, then the expected AIS is 4.

## 4. PHASE II: Accident reconstruction and AIS validation

## 4.1 Accident reconstruction

The accident data in this section were provided by the UKPF and the Senior Coroner and consisted of detailed Police accident as well as PM reports. The accident reconstruction focused on re-creating the vehicle and the accident circumstances. In order to capture the pedestrian kinematics, the THUMS human model was scaled to match the height and weight of the deceased, and placed in the most likely gait [15][16], based on of accident report, to replicate an accurate head landing position on the windscreen.

The cases studies are listed in Table 3.

| Case Id: | UK Police Force Reference | Vehicle | Mass (kg) | Height (cm) | Impact direction | Vehicle Impact Speed (m/s) |
|---|---|---|---|---|---|---|
| 1 | 229-4818 | Seat Leon | 61.0 | 183 | Left side impact (right leg forward) | 16.0 |
| 2 | 213-2205 | Toyota Corolla | 58.6 | 165 | Right side impact (right leg forward) | 11.2 |
| 3 | 001-3484 | Renault Clio | 79.2 | 173 | Side (left leg forward) | 12.5 |
| 4 | 207-9077 | Benz B180 | 56.4 | 165 | from driver's near to far side | 12.5 |

Table 3 Accident cases summary

Each vehicle stiffness was related to the EuroNCAP pedestrian scoring system [20][21][22][23] and the stiffness value characteristic inspired by the APROSYS project [25]. The method proposed varied in the way the contact characteristic was provided. In APROSYS Madymo was utilised, which favoured a contact force method which is well suited to multibody software. In the case of a full Finite Element model containing soft tissues, i.e. for which the stiffness values were very low compared to bones, similar contact method could not be achieved, as it gave model instability. Consequently, a penalty method was preferred by tuning the thickness of a bonnet area by simply changing the thickness whilst still meeting the EuroNCAP test results. The thickness of these panels is listed in Appendix D. The accident kinematics can be seen in Appendix E. It can be noted that the impact locations computed are similar to the real-life scenario (Figure 31). The accident kinematics are illustrated in Appendix E, in Figure 32, Figure 33, Figure 34 and Figure 35.

## 4.2 Traumatology results (numerical and real-life)

Results are plotted in Figure 8 to Figure 21. Black dots represent the CAE prediction results, while the red ones show the injury result based on the autopsy reports.

### 4.2.1 Case 1: 229-4818

In case 1, the PM listed that trauma was present on the right side of the brain. The pathologist did not give any information about brain contusion and corresponding side symptoms on the grey matter. In the CAE simulation, an AIS 2 brain contusion is observed (Figure 8). AIS 2 injury is a moderate injury which has 1%-2% probability of fatality [1]. Comparing with no injury suggested in the autopsy report, it can be noted that the CAE prediction is acceptable. It can be noted that in Figure 11, the trauma is more pronounced on the right-hand side, albeit small (AIS2). It can be observed that principal plastic strain response in Figure 11 is very scattered across the brain and does not show any clear trauma location, if compared to the PM. As a matter of fact, both the left and right side of the grey matter are injured, which is not what is expected.

On the white matter (Figure 9), due to a subdural hemorrhage identified in the PM report, an AIS 4 injury can be concluded. From CAE simulation, an AIS 2-3 DAI was confirmed (PVP values landing between AIS 2 and AIS 3 corridors). Trauma on the right-hand side of the brain is also observed on Figure 10 if PVP is used. It can be noted that there is a higher trauma in the center of the white matter, but such is not listed in the PM report. Looking at the principal plastic strains, the values are again scattered and do not suggest a clear trauma location. The difference in trauma results is due to the fact that the THUMS human computer cannot predict blood loss post-accident but only mechanical injury at the time of the accident, as such hemorrhage and swelling cannot be predicted using FEA. However, one of the side effects of DAI is hemorrhage, therefore corresponding hemorrhage can be assumed according to the AIS 3 prediction result.

Looking at the PVP values in the median area are higher than the right side of the brain. This has been missed in the PM. Nevertheless, the right impact location and trauma were predicted by the proposed method (right side of the brain). The PVP of the left-hand side of the brain is lower than the right-hand side of the brain, hence the trauma could have been missed in the PM. Looking at the principal plastic strains, the values are in excess of 100% which would suggest AIS 4 if not AIS 5, considering Table 1, which is contradictory to the PM outcome. In this case, maximum principal strain does not capture the location nor the trauma level.

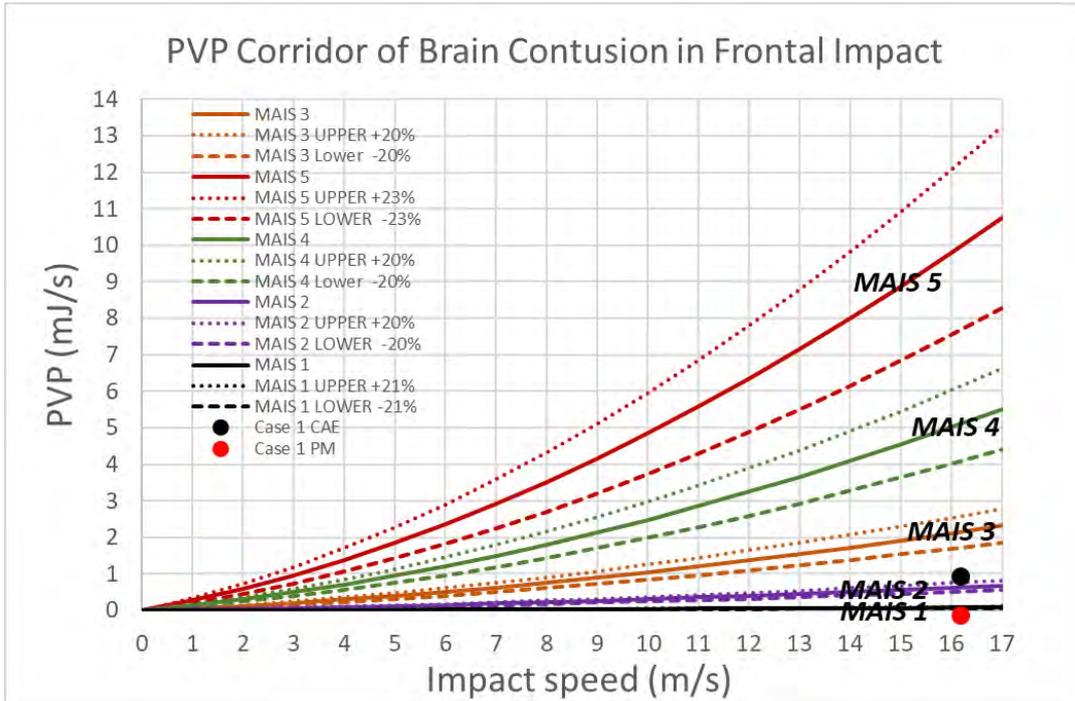

Figure 9: Case 1: Brain contusion result of case 1 from CAE and autopsy report (grey matter)

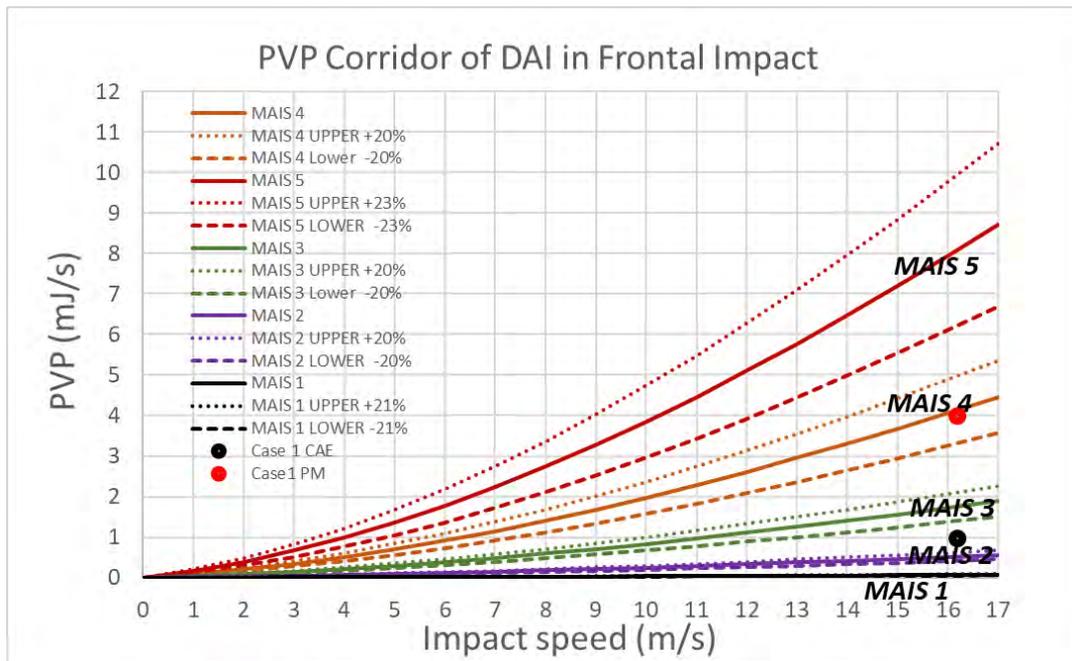

Figure 10: Case 1: DAI result of case 1 from CAE and autopsy report (white matter)

Comparing the location of the impact, PVP can be compared with the current method widely used which is the maximum principal strain (Figure 10). It can be noted that PVP computes a trauma location median with a slight bias to the right, while the plot with the maximum principal strain is not conclusive in location as well as in AIS outcome.

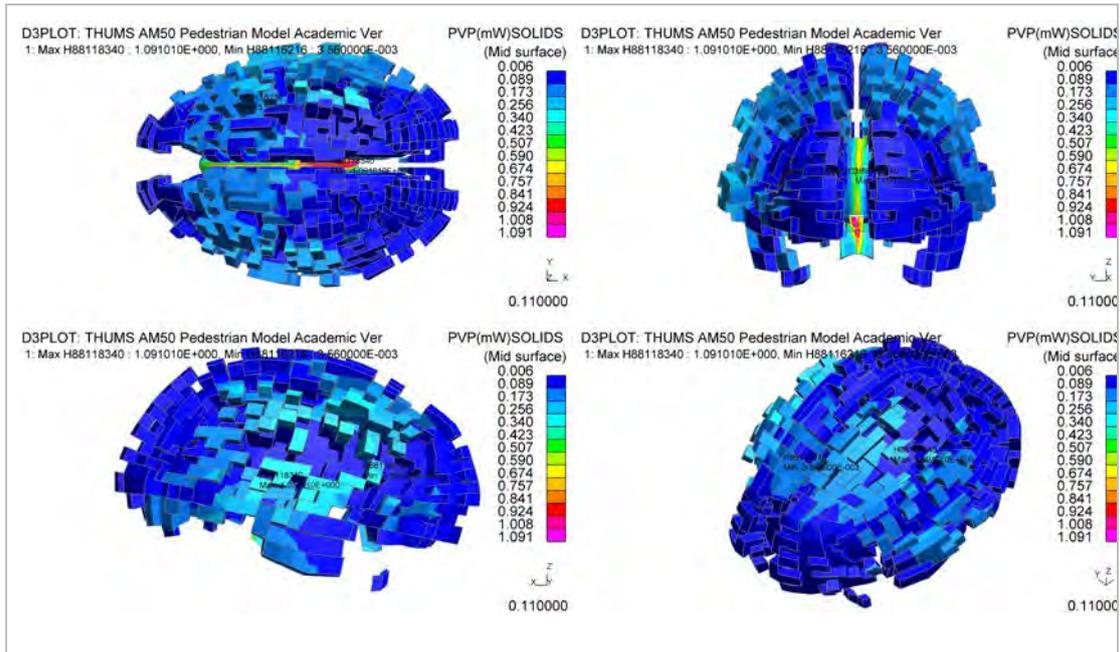

White Matter PVP results

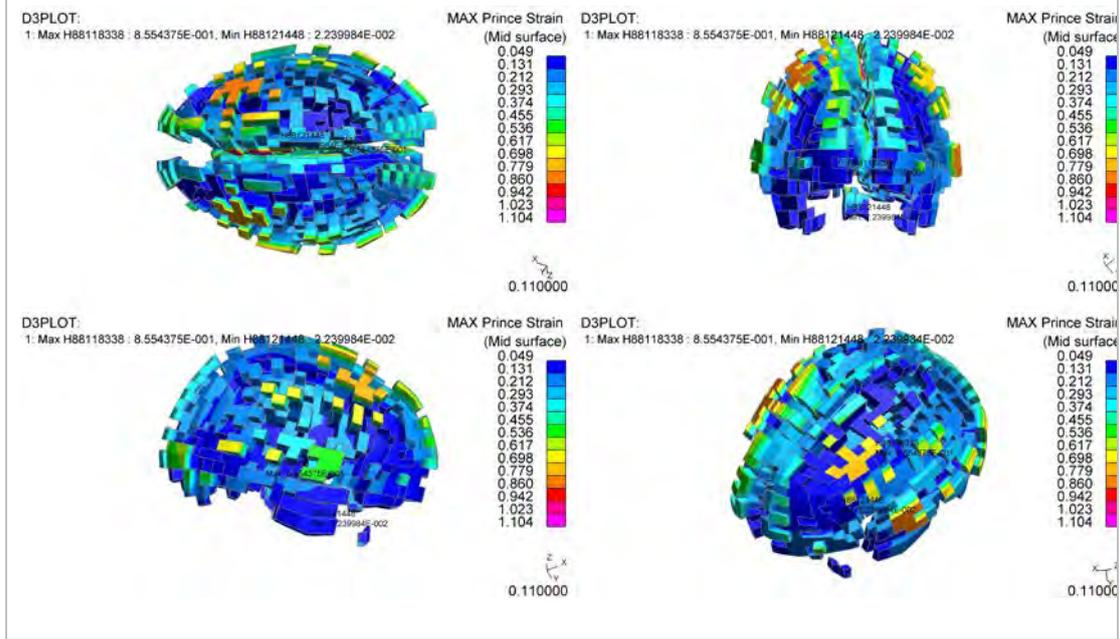

White Matter Principal Strain Results

Figure 11: Case1 - White Matter injury comparison between PVP and maximum principal strain

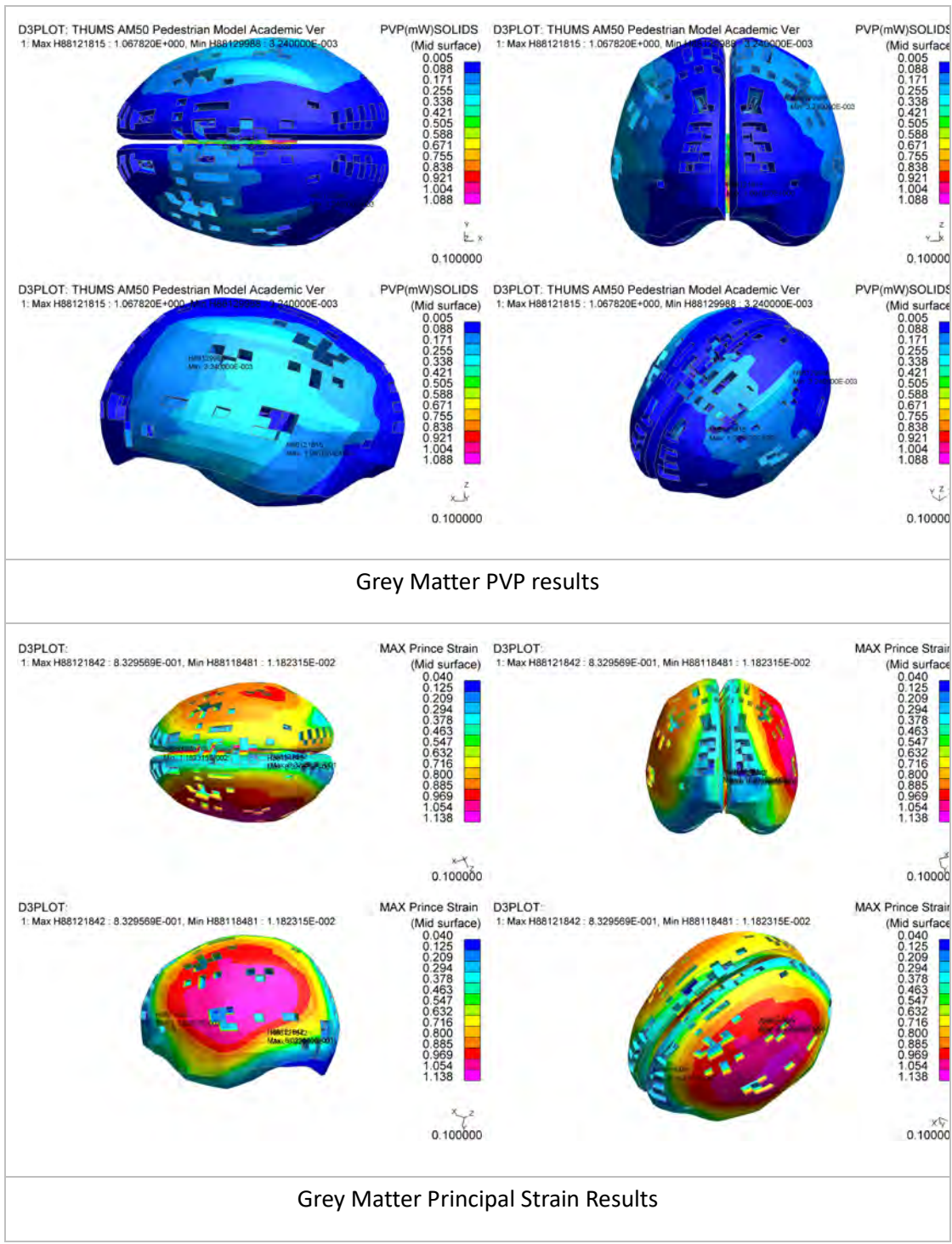

Figure 12: Case 1 - Grey Matter injury comparison between PVP and maximum principal strain

### 4.2.2   Case 2: 213-2205

In case 2, the PM listed that trauma was present on the right temporal lobe. Considering the brain grey matter (Figure 12), the PVP prediction and autopsy report are comparable. On the white matter (Figure 13), an AIS DAI injury can be concluded from both the PVP prediction and the autopsy report.

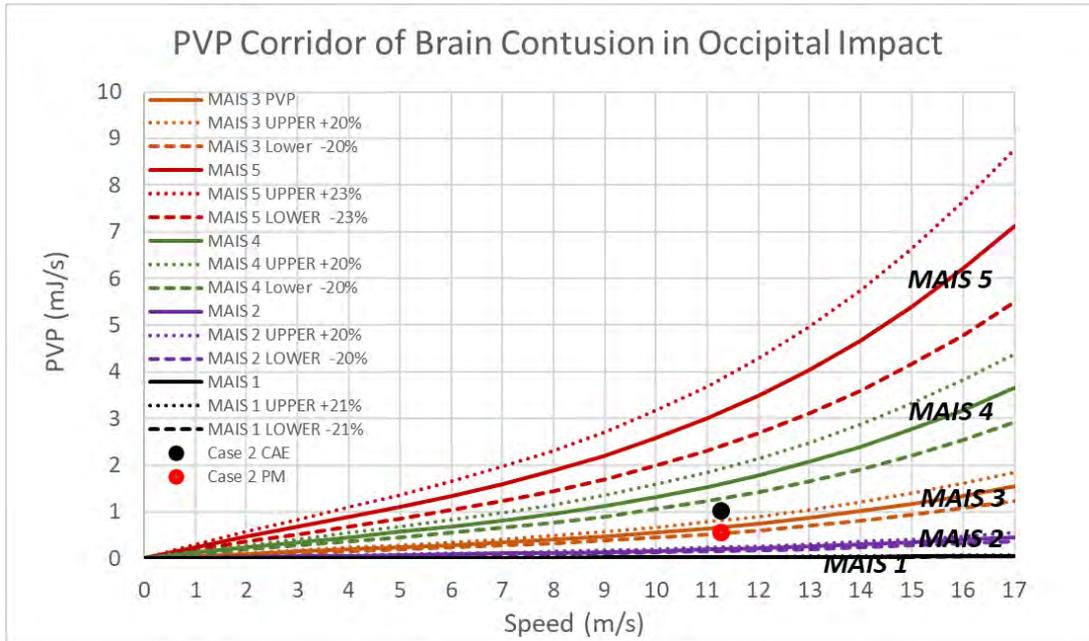

Figure 13: Case 2: Brain contusion result of case 2 from CAE and autopsy report (grey matter)

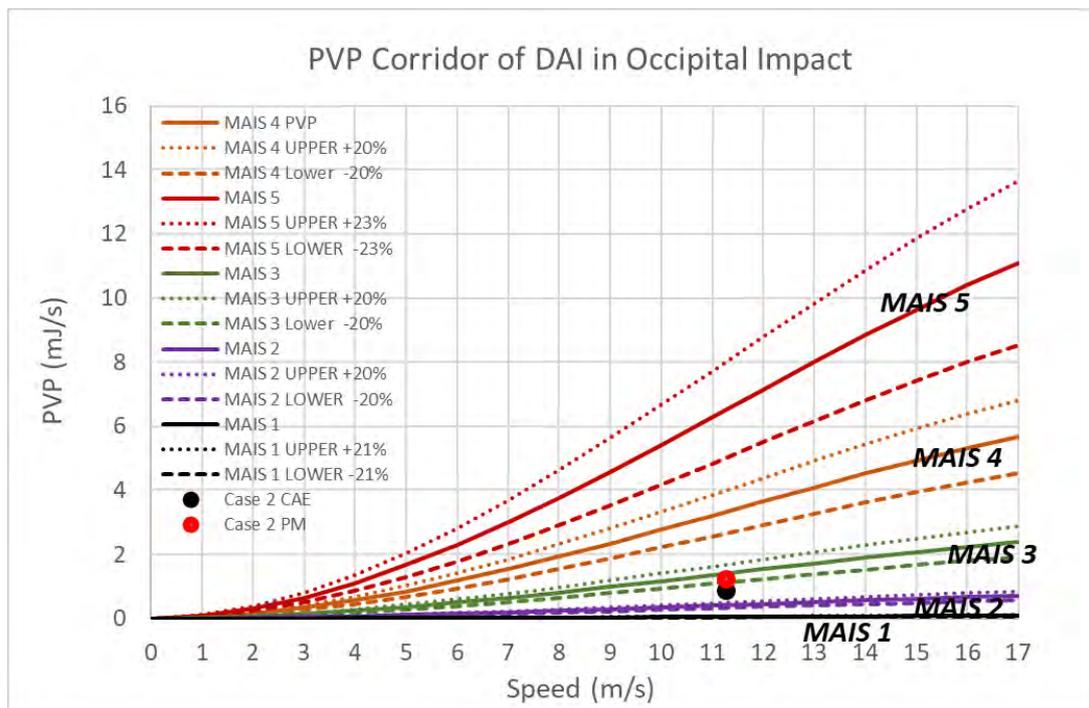

Figure 14: Case 2: DAI result of case 2 from CAE and autopsy report (white matter)

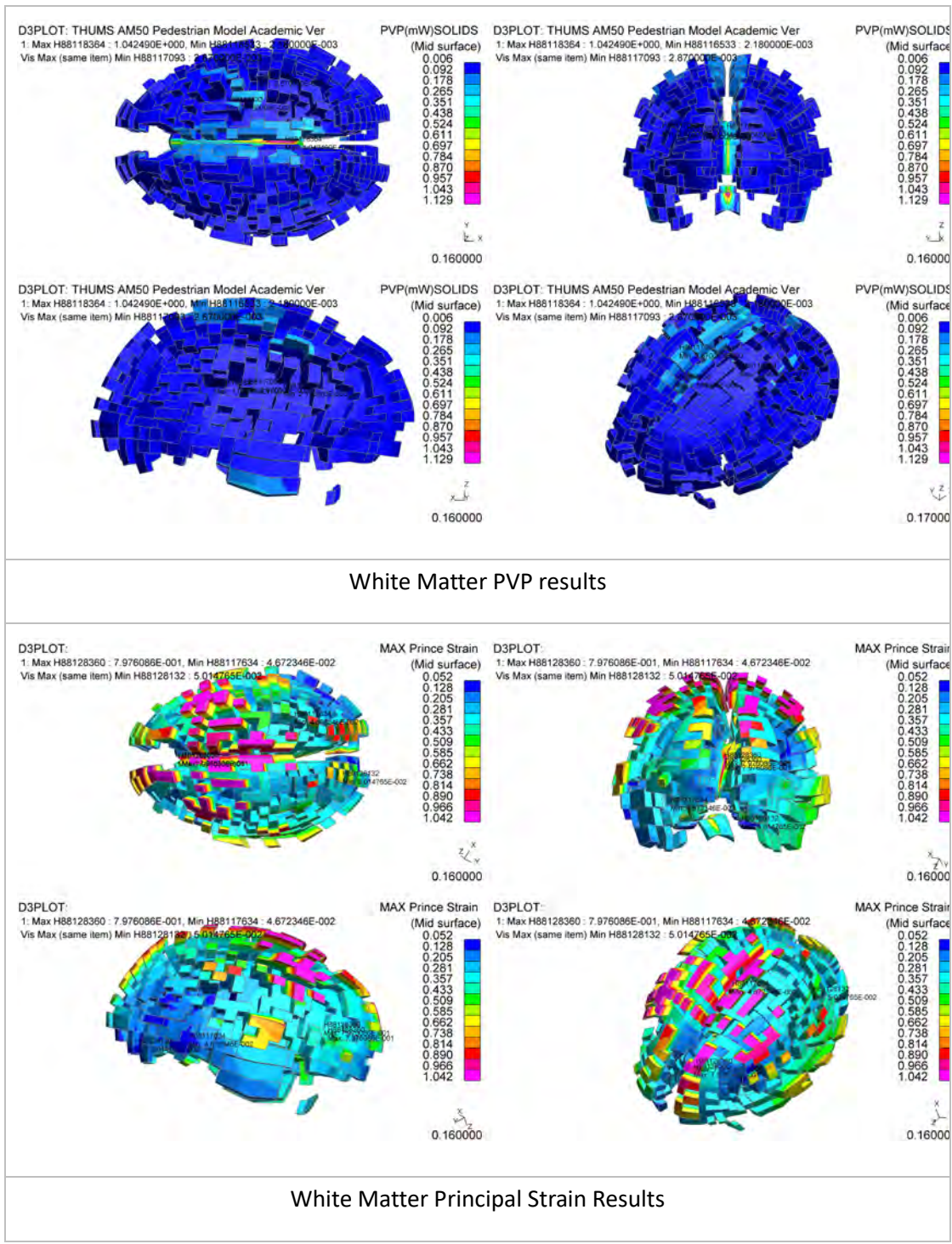

Figure 15: Case 2 - White Matter injury comparison between PVP and maximum principal strain

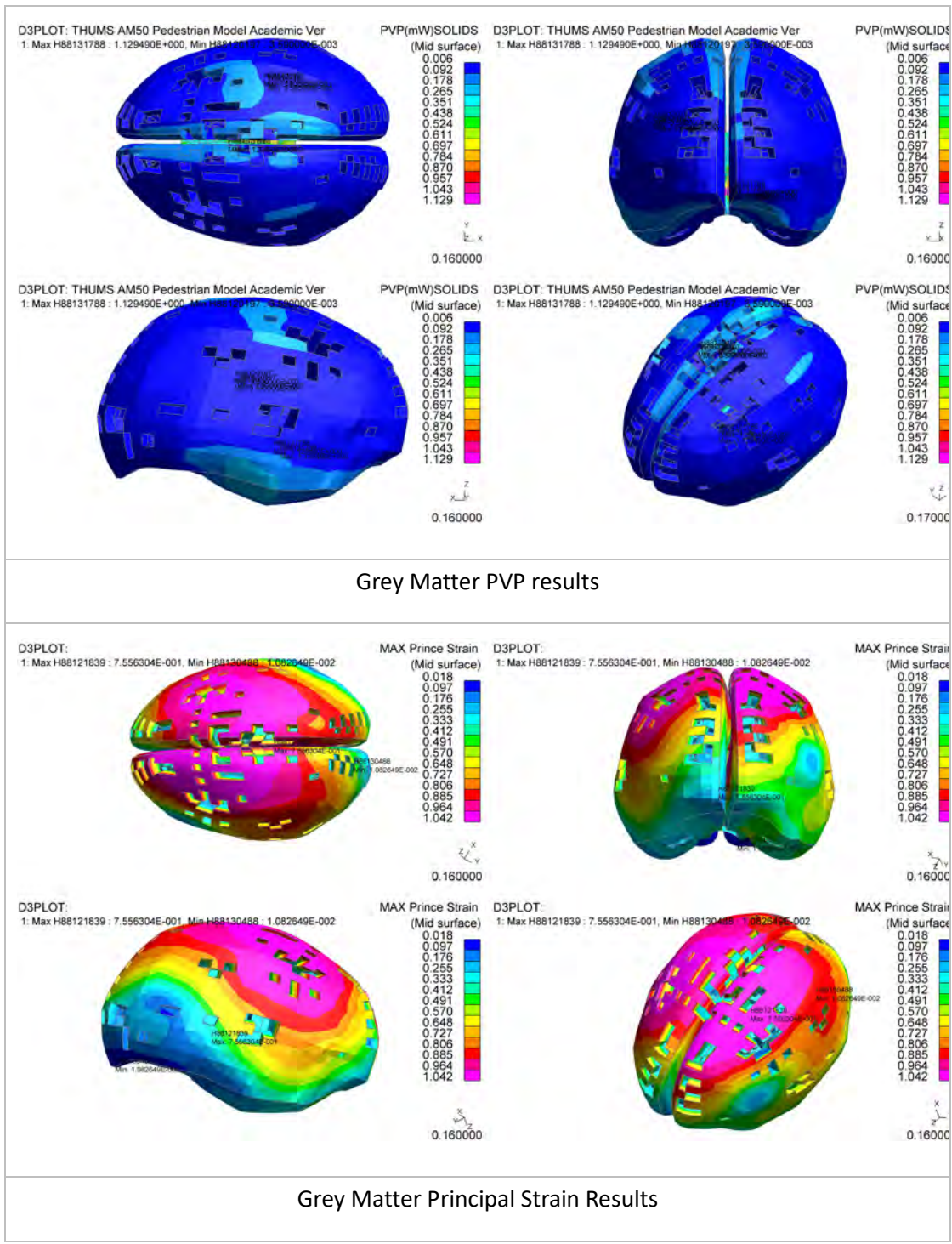

Figure 16: Case 2 - Grey Matter injury comparison between PVP and maximum principal strain

It can be observed from Figure 15 that the right temporal area has been injured, just by looking at the PVP plot. It is suggested that the parietal lobe would be also damaged with an AIS2, but was missed in the PM. When the principal strain plots are observed, they suggest that the values are high around the parietal area, which is in the wrong location and not in the temporal area; these strain values also tend to scatter as in Case 1. Looking at the principal plastic strains, the values are in excess of 100% which would suggest AIS 4 if not AIS 5, considering Table 1, which is contradictory to the PM outcome. In this case, maximum principal strain does not capture the location nor the trauma level.

### 4.2.3   Case 3: 001-3484

In case 3, the pathologist did not observe any injury on the brain tissue, while using PVP, AIS 3 brain injury can be concluded on the grey matter (Figure 16) and AIS2 on the white matter (Figure 17). No injury description was given in the autopsy report; however, the pedestrian death was recorded as death from multiple injuries. Therefore, based on the autopsy report, MAIS of the pedestrian should be 0 which does not correlate with real-life accident. The fact that no injuries were recorded in the PM does not mean that the injury was not present, but was probably too small to be observed by the pathologist. It can be suggested that PVP could suggest some trauma zones to the pathologist, like the upper lobes in this instance, which can be observed using the PVP output from Figure 18 and Figure 19.
Looking at Figure 18 and Figure 19, it can be noted that the PVP trauma plots are less scattered than using the standard maximum principal strain method. Also, in both cases, the maximum principal strain values are lower than the critical values from Table 1, but no AIS can be concluded from their values.

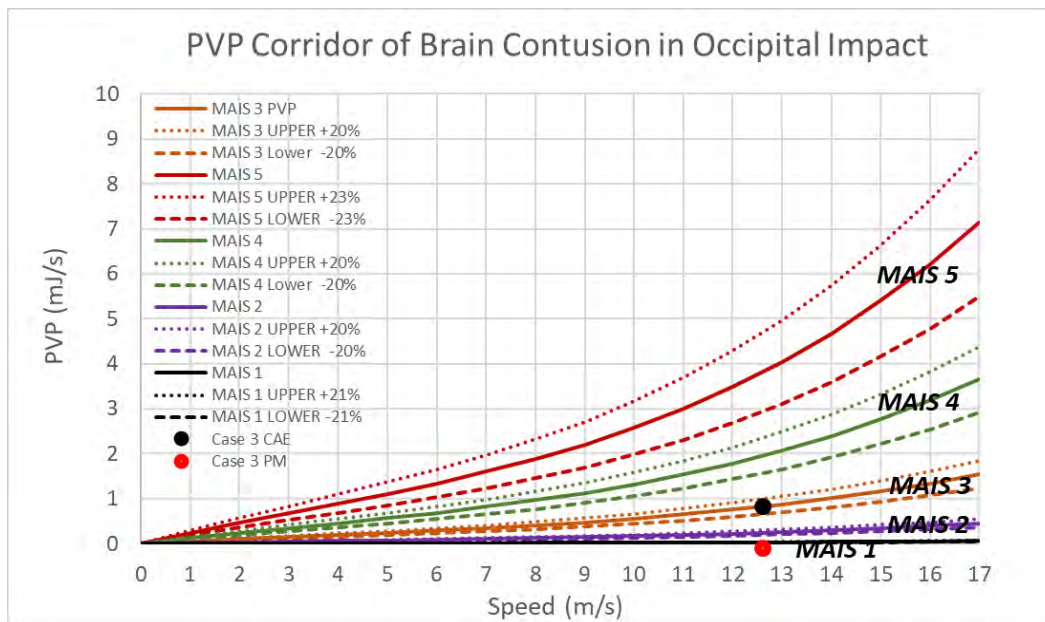

Figure 17: Case 3: Brain contusion result of case 3 from CAE and autopsy report Figure 17

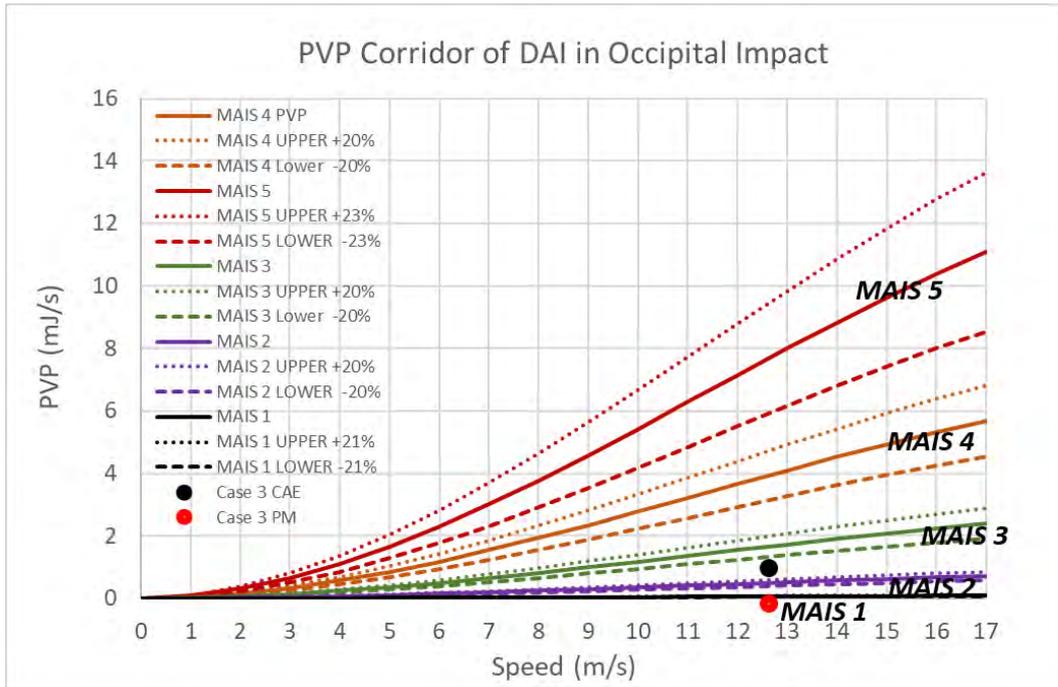

Figure 18: Case 3: DAI result of case 3 from CAE and autopsy report Figure 17

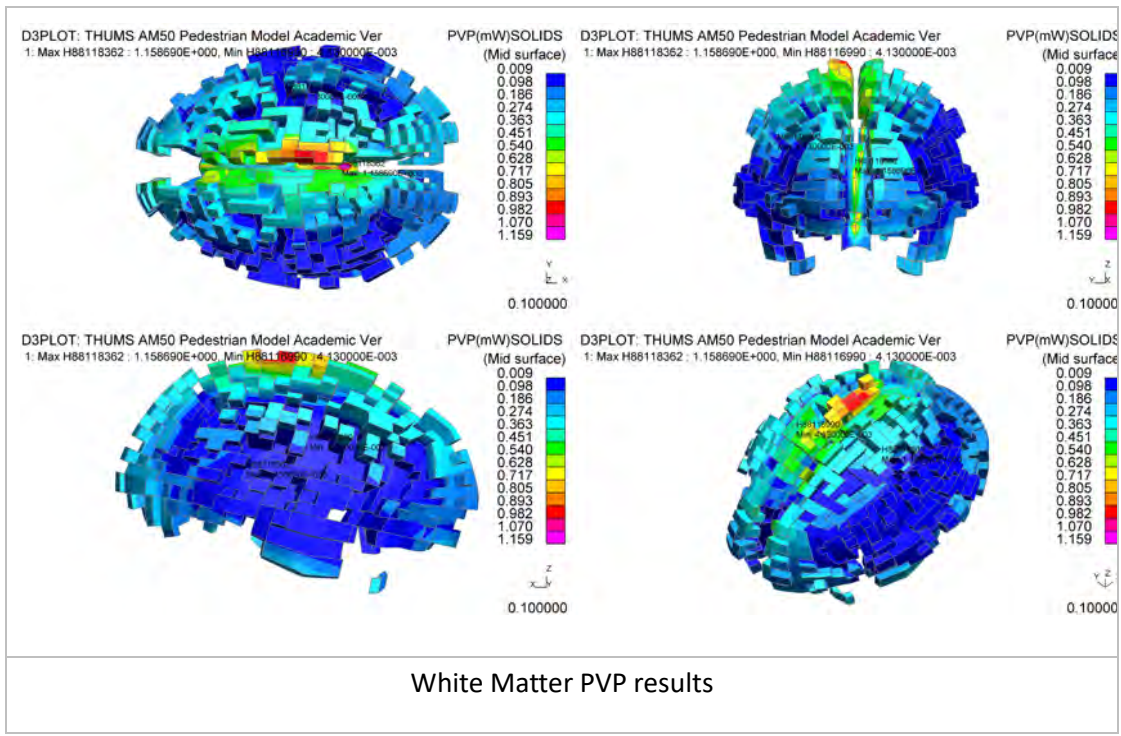

| White Matter PVP results |
|---|

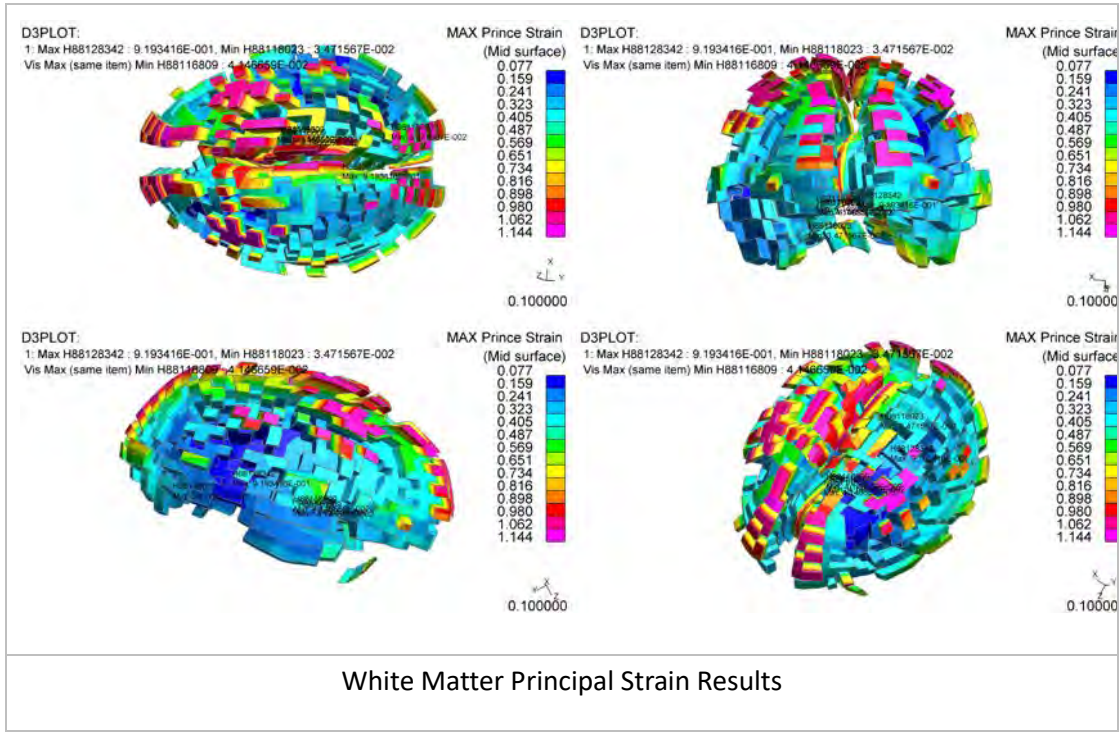

White Matter Principal Strain Results

Figure 19: Case 3 - White Matter injury comparison between PVP and maximum principal strain

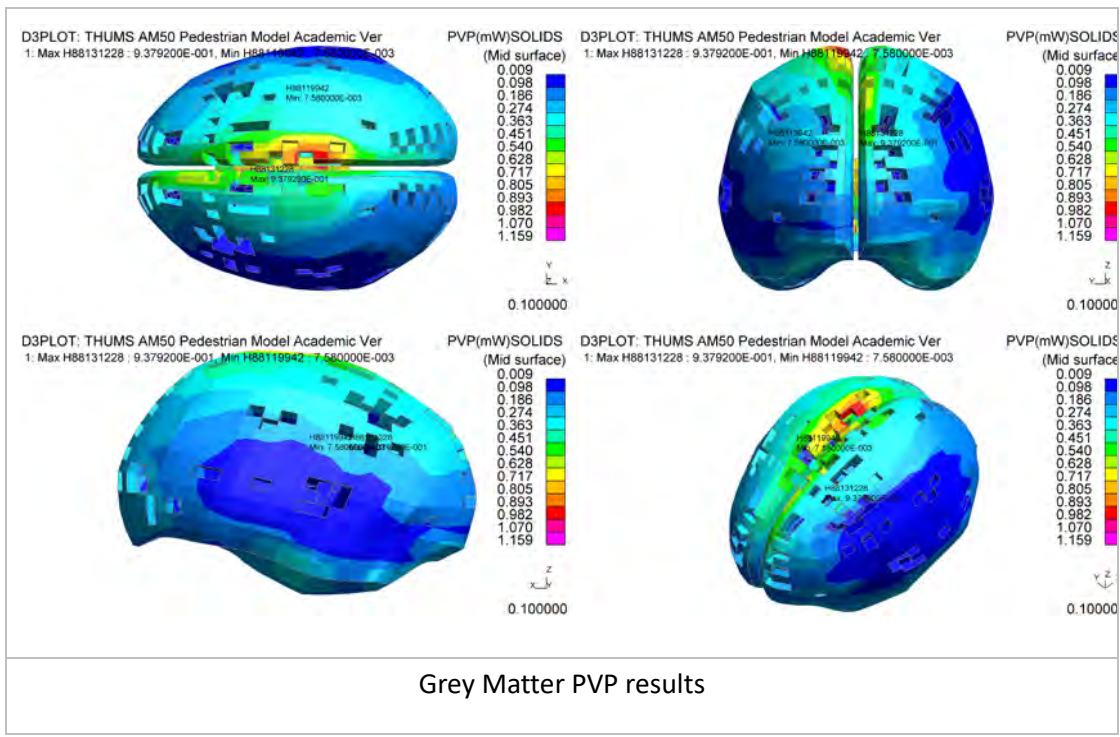

Grey Matter PVP results

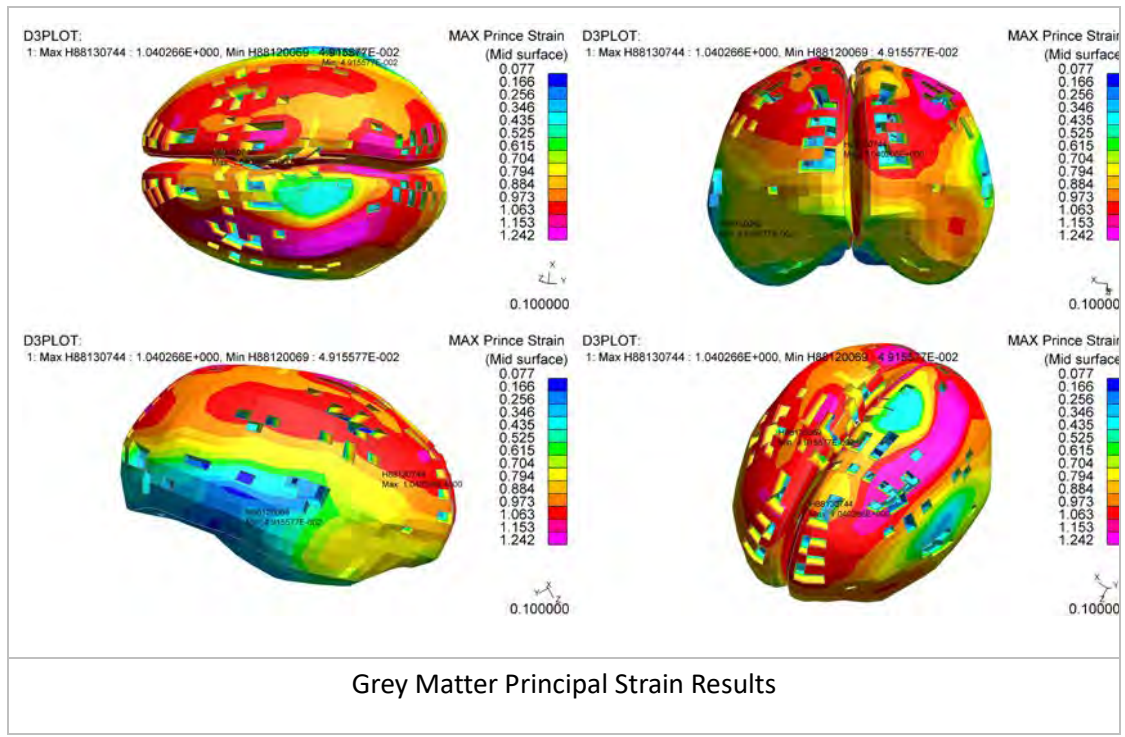

Grey Matter Principal Strain Results

Figure 20: Case 3 - Grey Matter injury comparison between PVP and maximum principal strain

### 4.2.4 Case 4: 207-9077

In case 4, the PM suggested extensive trauma on both lobes, which can be observed in the PVP plots from Figure 22 and Figure 23. No brain contusion was observed by the pathologist, and PVP suggests an AIS2 outcome (Figure 20), which is a reasonable match as AIS 2 may be too small to be observed during a PM. An AIS 4 caused by DAI was concluded in the autopsy report due to the subdural hemorrhage. On the white matter PVP predicted an AIS 3 (Figure 21). Again, due to the limitations, PVP/FEA cannot predict post-accident injury and so hemorrhaging is out of the capabilities of the PVP/FEA prediction. It can be noted that in the PVP plots of the white and grey matter, the maximum PVP appears in the median area of the brain. This was not captured in the PM. Regarding the maximum plastic strain, the same comment can be made, i.e. the location and the AIS predictions are not representative to what happened during the accident.

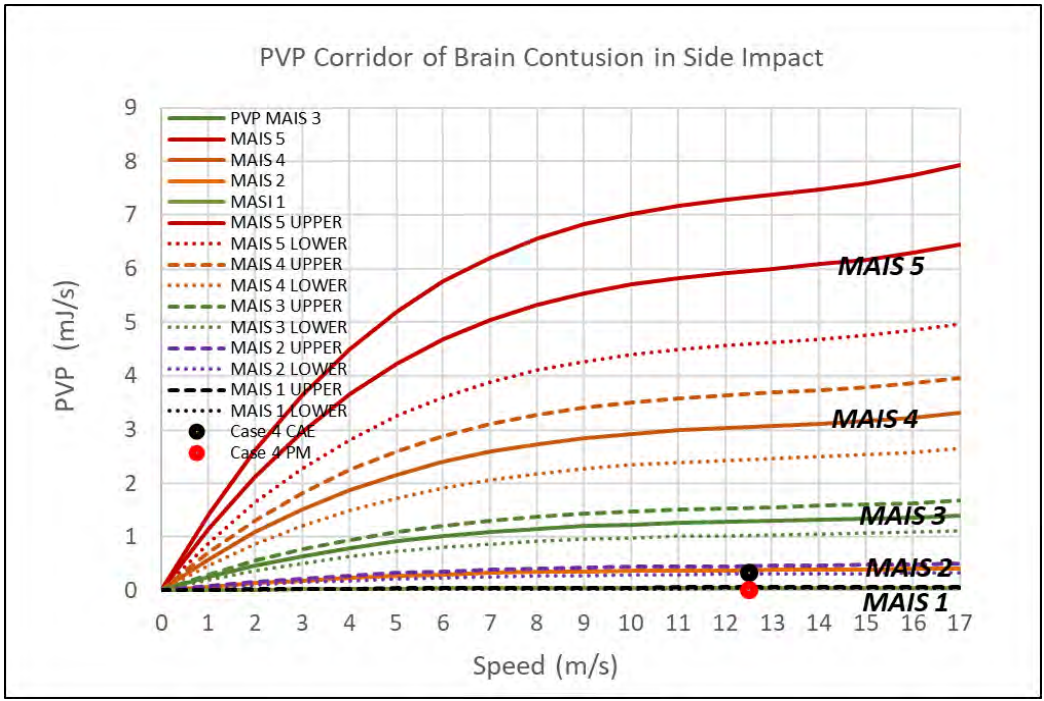

Figure 21: Case 4: Brain contusion result of case 4 from CAE and autopsy report Figure 17

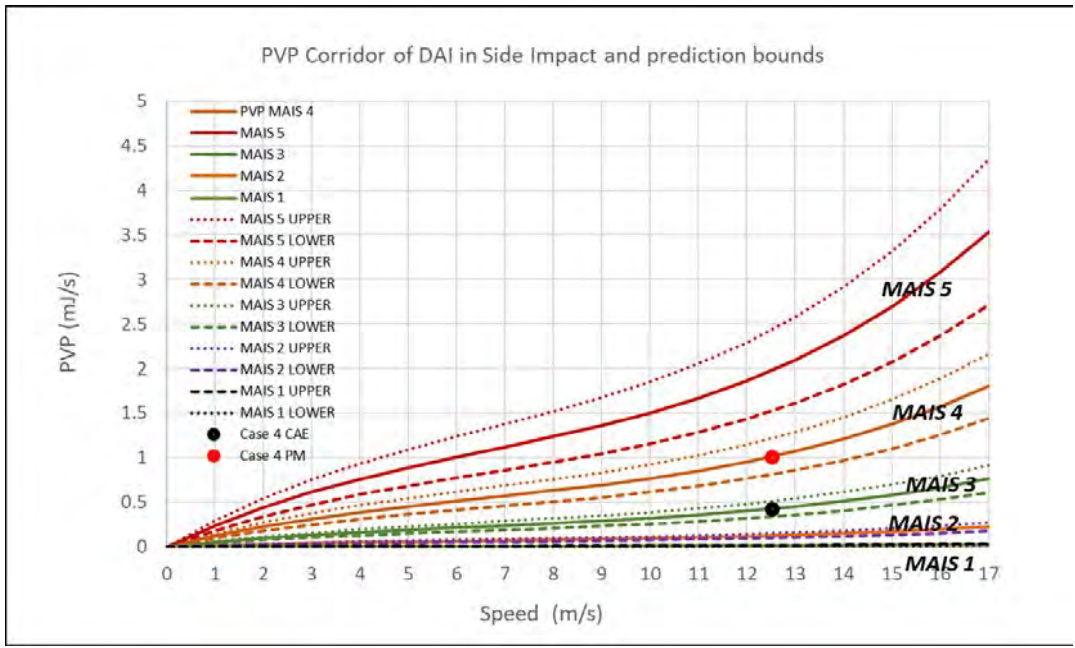

Figure 22: Case 4: DAI result of case 4 from CAE and autopsy report Figure 17

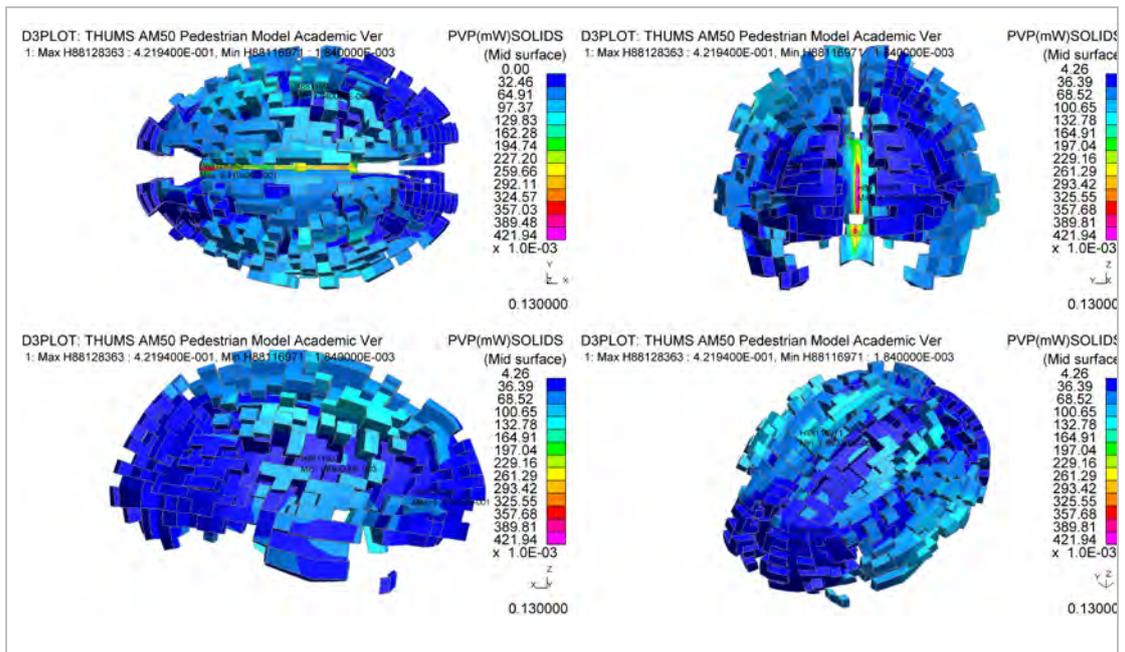

White Matter PVP results

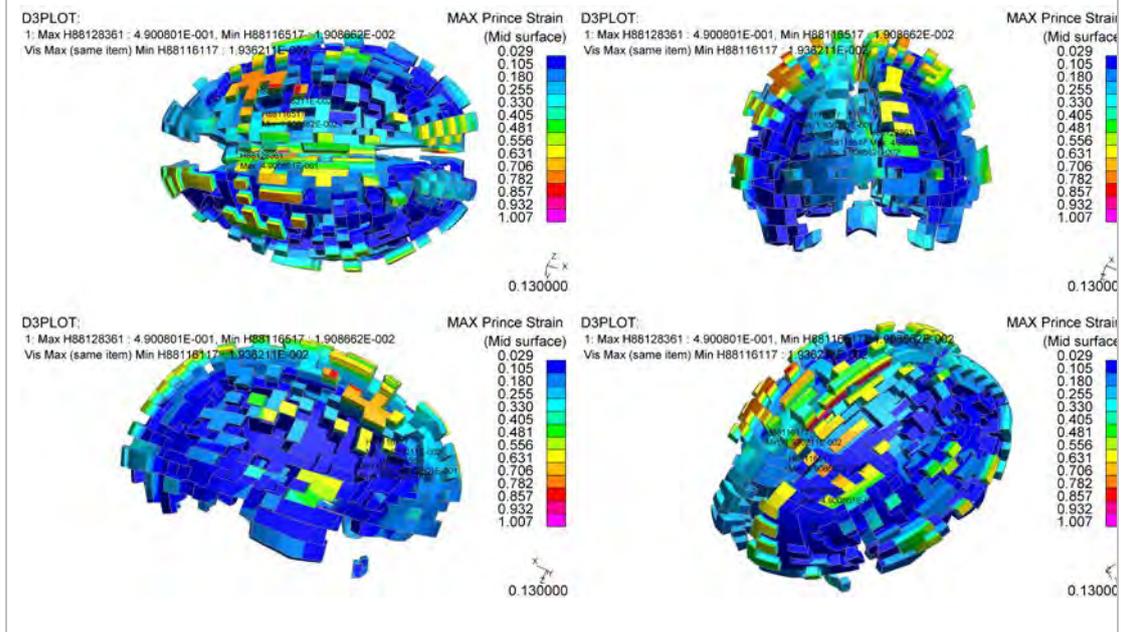

White Matter Principal Strain Results

Figure 23: Case 4 - White Matter injury comparison between PVP and maximum principal strain

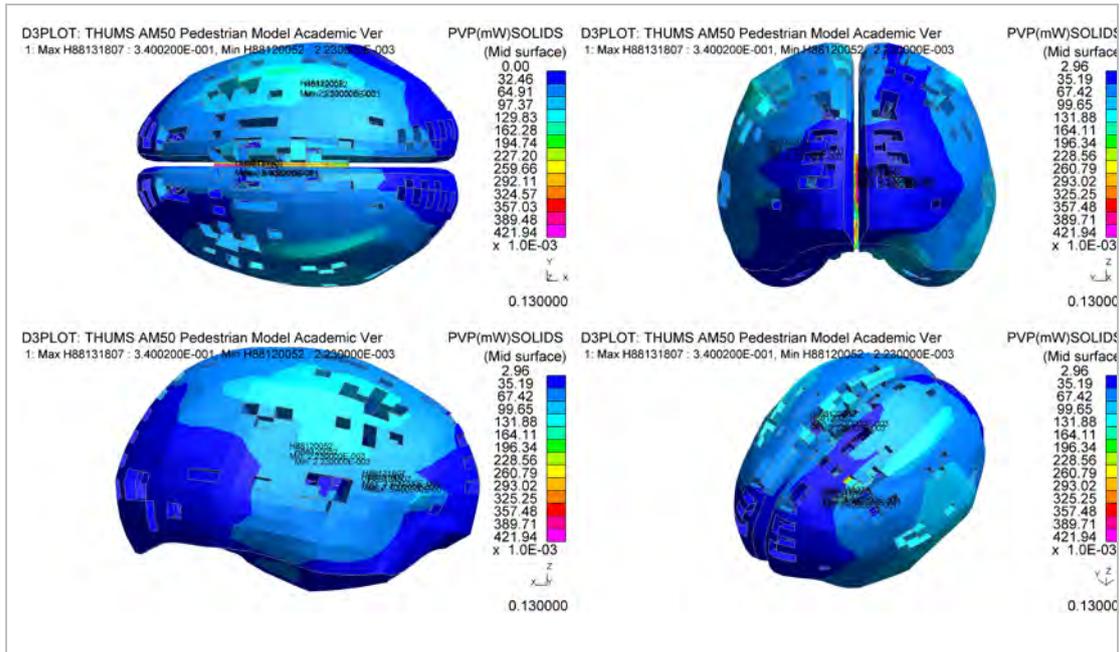

Grey Matter PVP results

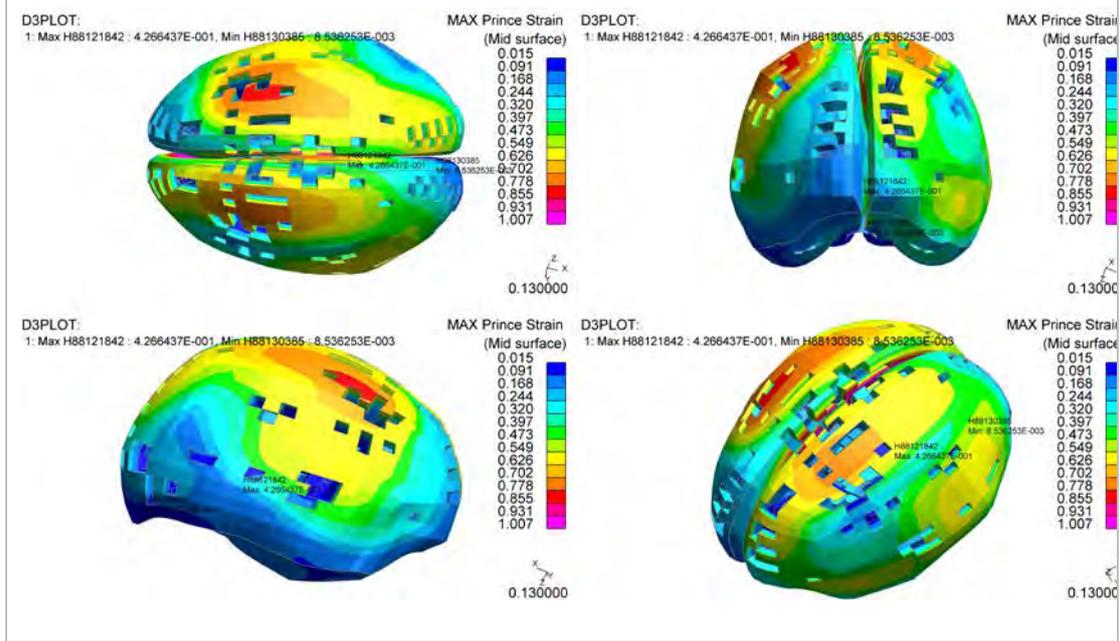

Grey Matter Principal Strain Results

Figure 24: Case 4 - Grey Matter injury comparison between PVP and maximum principal strain

# 5 Discussion

The results of the four accidents can be summarized in Table 4.

| Case Id: | Vehicle | PM description | CAE location | Comments | Match (Y/N) |
|---|---|---|---|---|---|
| 1 | Seat Leon | Right parietal lobe (Haemorrhage) | 88118340 (Centre of parietal lobe, not surface) | Haemorrhage is not of the prediction scope of THUMS. | Yes |
| 2 | Toyota Corolla | The inferior aspect of the right temporal lobe (Swollen) | 88131788 (Left grey matter, but in midline of whole grey matter) | The maximum PVP is not located on the description of PM, but PVP distribution and trauma level can be clearly observed on right temporal lobe. | Yes |
| 3 | Renault Clio | No evidence of skull fracture and brain showed no evidence of contusion | 88118362 (Centre of right white matter parietal lobe) 88121842 (Centre of right grey matter parietal) | Trauma is too small to be captured by a PM | No, but prediction plausible |
| 4 | Benz B180 | Extensive haemorrhage within the left cerebral hemisphere with peripheral haemorrhage within both cerebral lobes. | 88128363 (Left white matter frontal lobe) | The maximum PVP is not located on the description of PM, but PVP distribution and trauma level can be clearly observed on both lobes | Yes |

Table 4: Summary of the accident study

In the first instance, it can be observed that the CAE predictions are of the same order of magnitude as the PM's. In the instance of Case 1, the CAE prediction was AIS 3 while the PM predicted AIS 4. In the PM report, it was observed that the skull and the cranial cavity were normal. The brain showed an area of subdural haemorrhage over the right parietal lobe, and also over the cerebellum in the midline and over right cerebella hemisphere. The cut surface of brain showed some small petechial haemorrhage present in right cerebella peduncle. No other brain injury was identified. In Case 1 it can be observed from Table 6 (Appendix F) that the PVP is located in the exact area of the PM, albeit right from the brain centreline. The pedestrian died, however, of a brain haemorrhage. The OTM model is based on a CAE model which can only predict mechanical damage, and not the blood loss, which is a fluid problem. Nevertheless, as a mechanical indicator it predicted the correct damage area.

In Case 2, CAE and PM both predicted an AIS of 3 on the brain, which is a serious injury. The PM listed that there was some subarachnoid haemorrhage. The brain appeared diffusely swollen to a mild degree and there were contusions on the inferior aspect of the right temporal lobe. These contusions were captured, however it was not possible to predict the haemorrhage and the swellings which are occurring post impact.

Case 3's PM was unremarkable, as no trauma was observed in the deceased ("No evidence of skull fracture and brain showed no evidence of contusion"). The CAE model predicted an AIS of 2 which is a moderate trauma. Maybe such trauma level is quite complicated to observe, as being low risk, hence it may be suggested that the CAE prediction is plausible.

Case 4 had some similarity with Case 2, except that there was some "significant" skull fracture, which had not been activated during the computations. This fracture is extending from the right temporal area coronally to the left temporal region. There was also a fracture of the base of skull on the left-hand side. A subarachnoid haemorrhage was identified and, on serial slicing through the brain, there was extensive haemorrhage within the left cerebral hemisphere with peripheral haemorrhage within both cerebral lobes. A 1cm haematoma was also noted in the right cerebellum. The computer model predicted an AIS 3 while the PM suggested an AIS 4. Again, it was not possible to predict the haemorrhage which is a post trauma effect which requires an Eulerian solving method to extract.

It can be noted that the UKPF is using the pedestrian kinematic effects to evaluate the vehicle impact speed, but not the Post-Mortem (PM), which contains vital information on the impact energy that was exerted to kill the pedestrian. It is not usually used since evidence from the PM would need to be presented by an expert (Home Office Forensic Pathologist). Overall the quality of autopsy reports (PM) is always questioned: just over half of PM reports (52%) (873/1,691) were considered satisfactory by experts, 19% (315/1,691) were good and 4% (67/1,691) were excellent. Over a quarter were marked as poor or unacceptable.

Proportionately, there were more reports rated 'unacceptable' for those cases that were performed in a local authority mortuary (21/214 for local authority mortuary cases versus 42/1,477 for hospital mortuary cases)" [19]. To date, experts tend to consider the research around PM currently to be limited regarding its use to predict speed, therefore would not use it in court during criminal proceedings. The PM is only used to state which organs failed, hence causing death, but not as evidence to add to the forensic case.

Overall, it can be observed that the comparison between PVP prediction result of pedestrian injury and autopsy report shows a promising correlation to risk to life applied to the head, in the trauma magnitude and location. This observation would suggest that it is possible to supplement the standard pedestrian head impactor numerical process with a human computer head model to assess the real trauma level of a pedestrian. With the current safety assessment processes which are using a head impactor, in ECE 127 [26] and EuroNCAP [27], which just evaluate HIC, it is only possible to evaluate the likelihood of linear skull fracture damage. This new method can go beyond the current limitations and predict the trauma outcome in the head's white and grey matter. In the case of EuroNCAP, as the bonnets are validated using calibration tests and then scaling of CAE prediction results, it would be possible to add this PVP method to simulate alongside the EuroNCAP protocol in order to supplement engineering assessment of brain injuries. Obviously, this suggestion would require a detailed test configuration setup, as the human head's trauma response is direction dependent; the current test assessment is made of half a sphere of rubber coated, aluminum consequently the proposed CAE assessment would require testing the 3 head directions studied in this report.

# 6 Conclusions

An Organ Trauma Model (OTM), based on Peak Virtual Power (PVP), was used to successfully extract the AIS risk to life, using the Finite Element method, to pedestrian white and grey matters in vehicle collisions. The OTM predicted trauma location as well as intensity, unlike current computer methods utilized. The OTM firstly calibrates PVP against the medical critical AIS threshold observed in each part of the head as a function of speed. This base PVP critical trauma function is then scaled and banded across all AIS levels using the property that AIS and the probability of death is statistically and numerically a cubic. The OTM was tested against four real-life accident scenarios for which PM data was available. The study concluded that PVP was a good candidate to predict AIS in a Finite Element head model, and that head trauma under-predictions were due to haemorrhage, which is post-impact. This method, however, brings some benefits, as it allows the assessment of head white and grey matter injuries, which are currently not measured, and may live alongside the current EuroNCAP test protocol to enhance the protection of pedestrian head injuries.

# 7 Limitations and Further Work

- THUMS is a dynamic Lagrangian CAE model which cannot be used to predict post-accident effects like swelling and bleeding, but the material damage: in this case trauma. Consequently, a means to extract the post-impact trauma will require a fundamental rebuild of the computer model and include maybe SPH or ALE formulations to evaluate bleeding and swelling.
- In the future, this study will continue and focus on other internal organs, like the liver, heart and kidneys, and investigate whether the same level of correlation can be achieved, leading eventually to the CAE calculation of the Injury Severity Score (ISS).
- As PVP is material property dependent, it would be theoretically possible to calibrate the OTM model with material properties for older people (Young's Modulus and failure strain level), making the OTM method a universal trauma modeling method.
- It could be hypothesized that pre-existing medical conditions could be pre-stored as a PVP value which could be added to the PVP generated by the collision.
- In the future, the OTM should be able to model and consider also failure, so that, for example, a broken rib could pierce a lung. Maybe PVP could be also part of a fracture model.

# Acknowledgements

The authors would like to thank the UK Police Force and the Coroners who made this research possible, as well as Dr Michal Orlowski and Mr Rohit Kshirsagar who have supported the authors with some of the CAE simulations and the post-processing of some trauma injuries.

# Appendix A: AIS in Lateral Head Impact

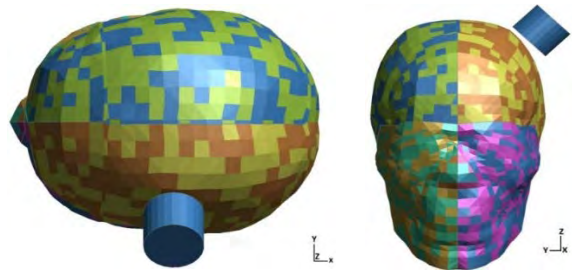

Figure 25 Scenario of parietal impact on THUMS' head

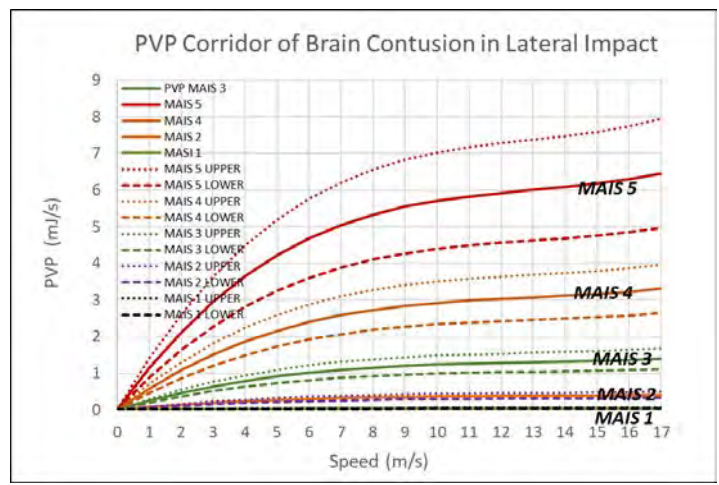

Figure 26 PVP corridor of brain contusion in lateral head impact

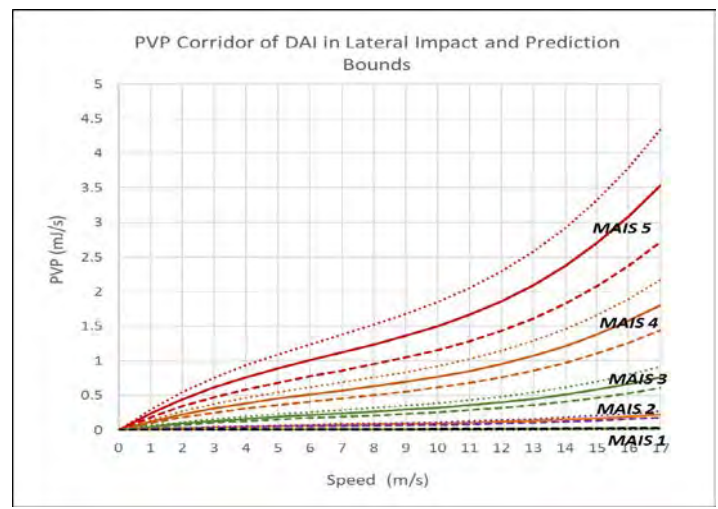

Figure 27 PVP corridor of DAI in head lateral impact

# Appendix B: AIS in Occipital Head Impact

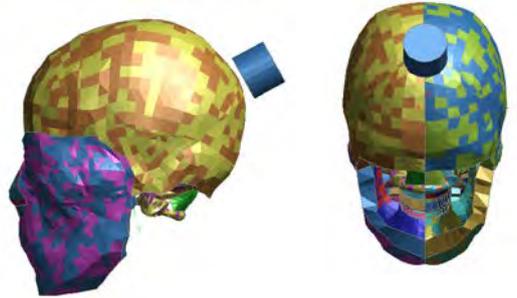

Figure 28 Scenario of Occipital impact on THUMS' head

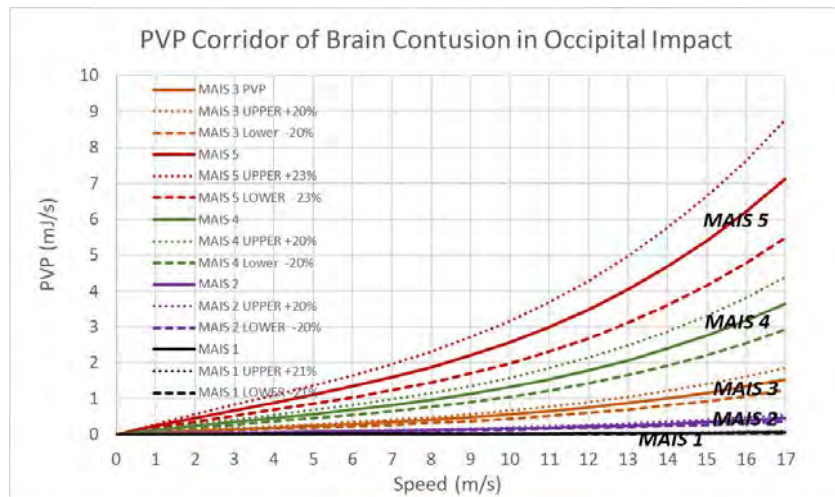

Figure 29 PVP corridor of brain contusion in occipital impact

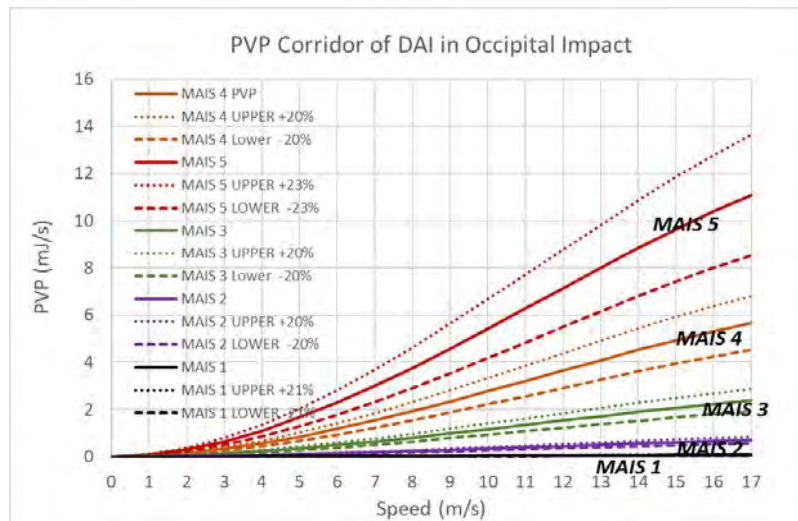

Figure 30 PVP corridor of DAI in occipital impact

# Appendix C: Stiffness map of each vehicle

| Case Id | EuroNCAP scoring | CAE model colour code |
|---|---|---|
| 1 [20] | 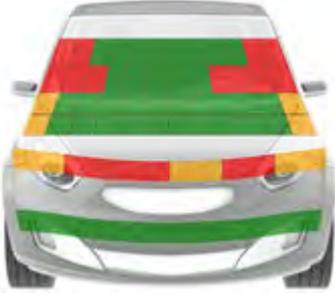 | 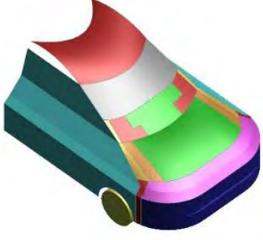 |
| 2 [21] | 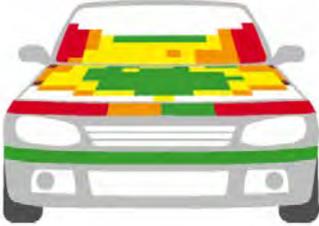 | 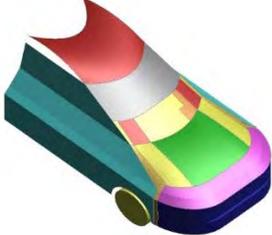 |
| 3 [22] | 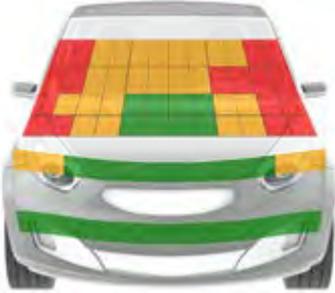 | 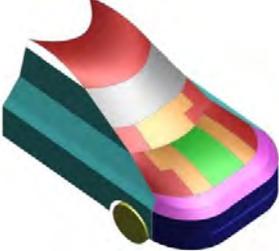 |
| 4 [23] | 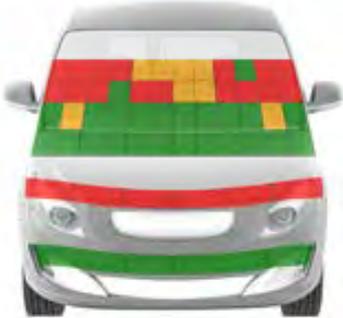 | 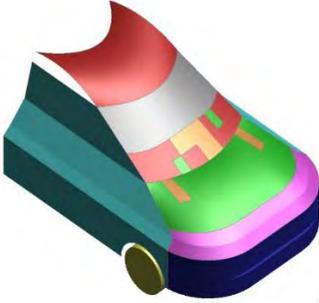 |

# Appendix D: Stiffness characteristic vs EuroNCAP map

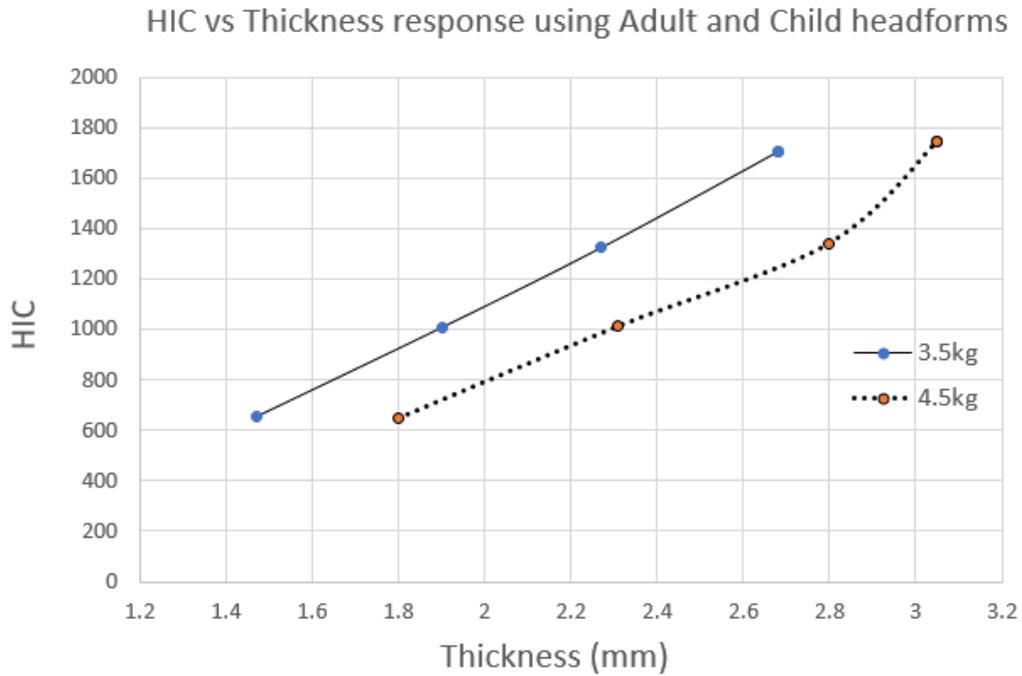

Figure 31 HIC vs thickness of 3.5kg and 4.5kg headforms

| Average HIC value as per EuroNCAP scheme | Impactor Mass | |
|---|---|---|
| | 3.5kg | 4.5kg |
| | Panel thickness (mm) | Panel thickness (mm) |
| Green (<650) | 1.47 | 1.80 |
| Yellow (650 – 1000) | 1.69 | 2.05 |
| Orange (1000 – 1350) | 2.09 | 2.56 |
| Brown (1350 – 1700) | 2.50 | 2.93 |
| Red (>1700) | 2.68 | 3.05 |

Table 5 Average HIC value using 3.5kg and 4.5kg headforms

# Appendix E: Accident Kinematics (1/3)

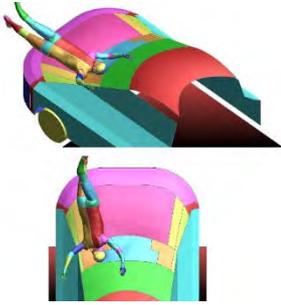

Figure 32: Validation of head impact location for each four accidents

# Appendix E: Accident Kinematics (2/3)

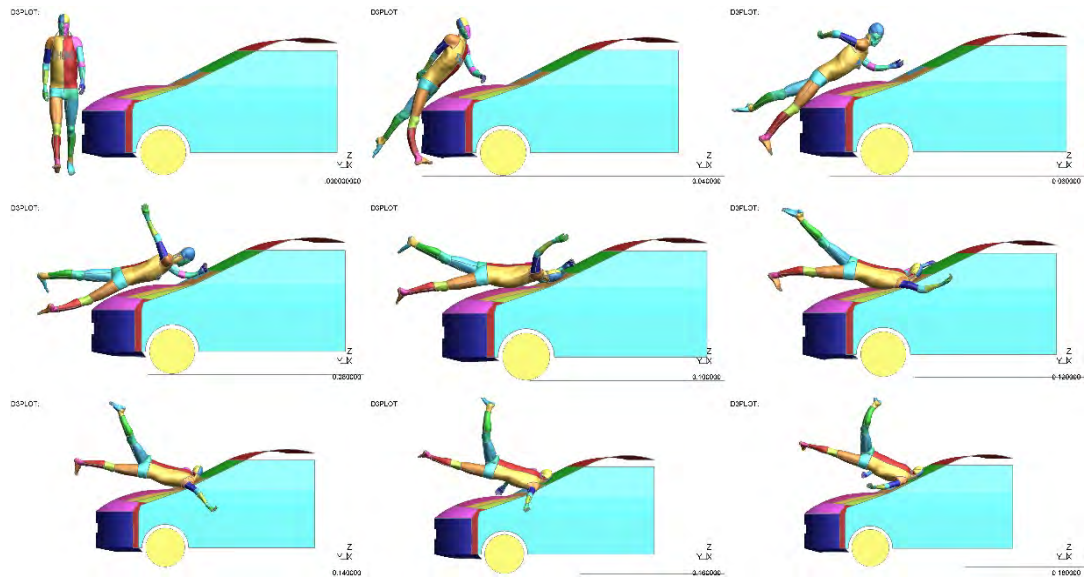

Figure 33: Seat Accident

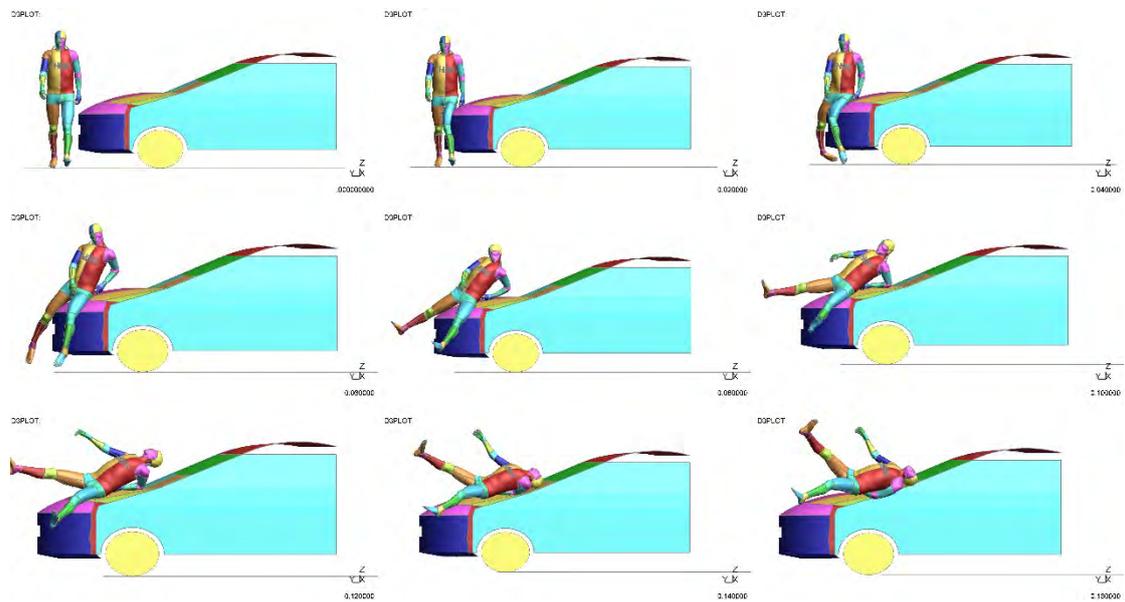

Figure 34: Toyota Accident

# Appendix E: Accident Kinematics (3/3)

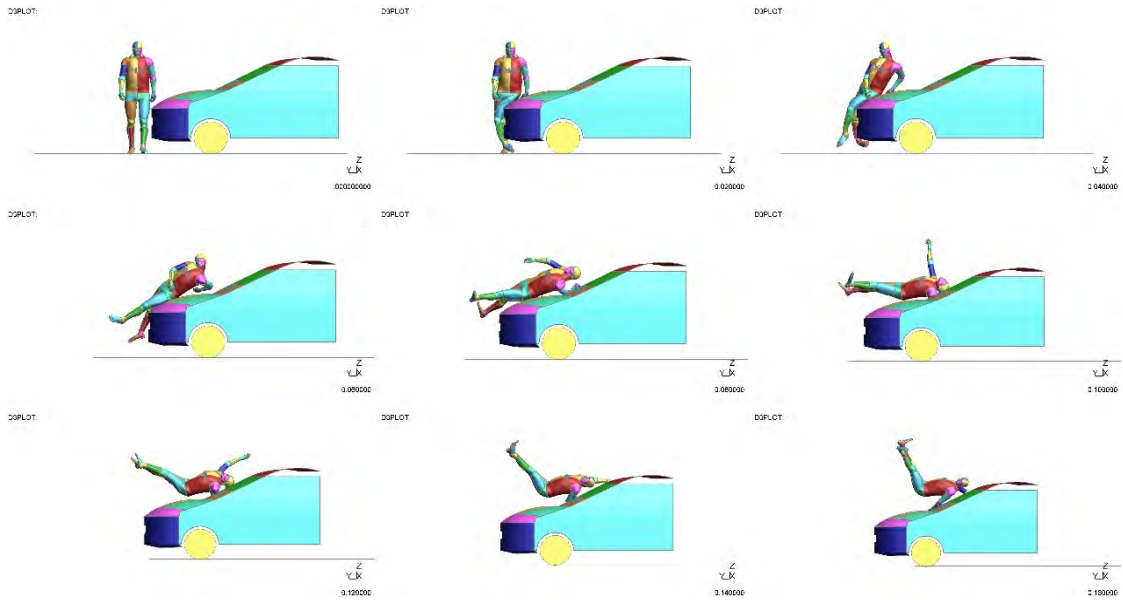

Figure 35: Renault Clio accident

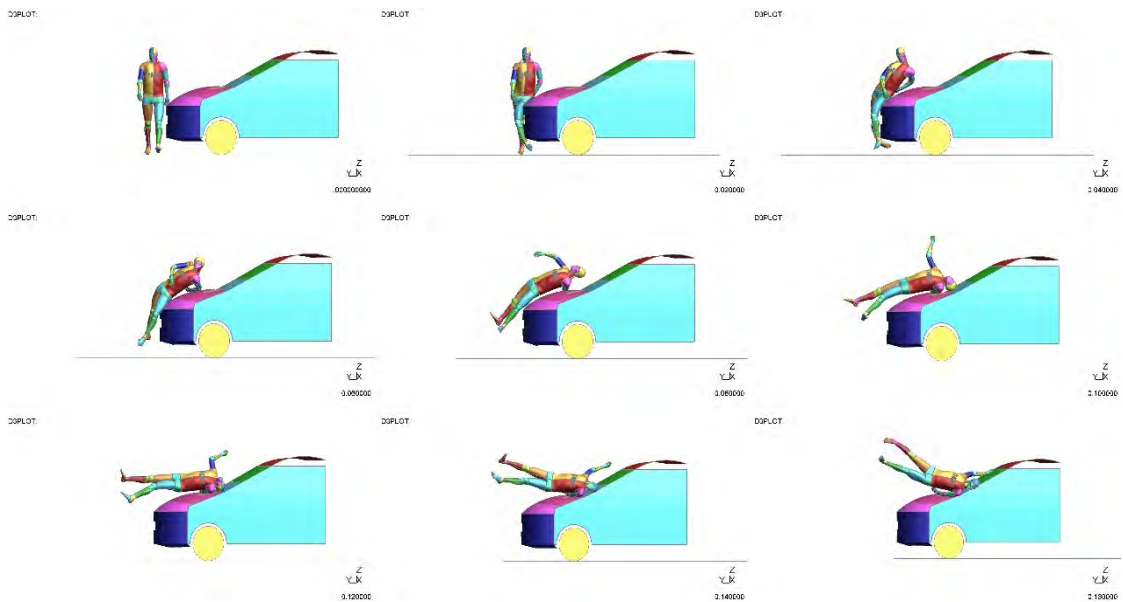

Figure 36: Benz B180 Accident

# Appendix F: Head Trauma location and PVP value (1/2)

| CASE Id and organ (white/grey matter) | PVP value from CAE (mJ/s) | PVP calibration value (mJ/s) | Impact mode (frontal, lateral, occipital) | Left / Right | Location within the brain |
|---|---|---|---|---|---|
| Case 1 (grey matter) | 1.07 | 0.00 | Frontal | Right | 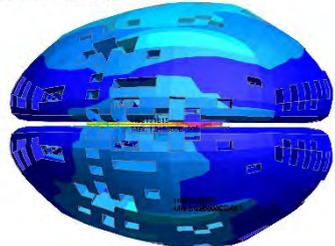 |
| Case 1 (white matter) | 1.09 | 4.10 | Frontal | Right | 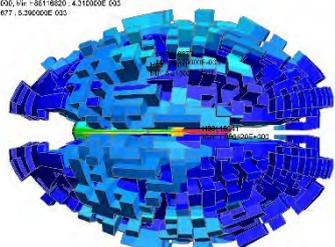 |
| Case 2 (grey matter) | 1.13 | 0.67 | Occipital | Right | 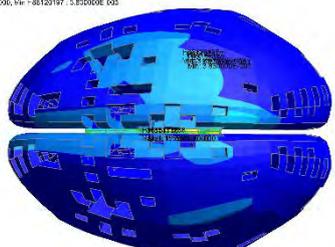 |
| Case 2 (white matter) | 1.04 | 1.39 | Occipital | Right | 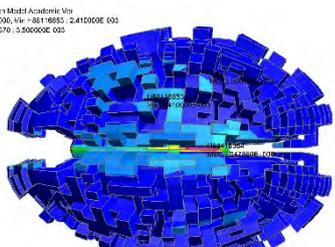 |

# Appendix F: Head Trauma location and PVP value (2/2)

| | | | | | |
|---|---|---|---|---|---|
| Case 3 (grey matter) | 0.94 | 0.00 | Occipital | Right | 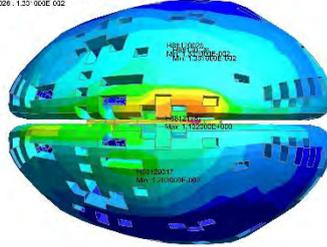 |
| Case 3 (white matter) | 1.16 | 0.00 | Occipital | Right | 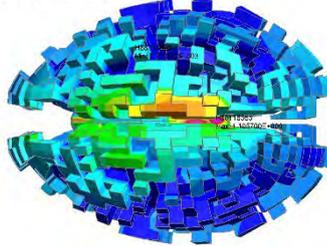 |
| Case 4 (grey matter) | 0.34 | 0.00 | Lateral | Both | 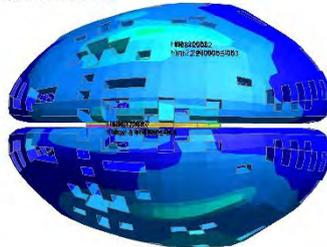 |
| Accident 4 white matter | 0.42 | 1.01 | Lateral | Both | 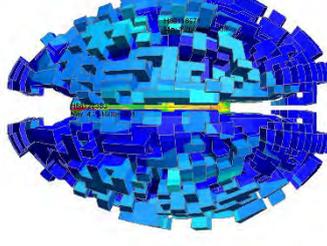 |

Table 6 PVP value and location of CAE result